\documentclass[12pt]{amsart}

\usepackage{amsmath, amssymb, graphics}
\usepackage{graphicx}

\input{epsf}

\newtheorem{thm}{Theorem}[section]
\newtheorem{cor}[thm]{Corollary}
\newtheorem{lem}[thm]{Lemma}

\newtheorem{defn}[thm]{Definition}

\numberwithin{equation}{section}

\newcommand{\eps}{\varepsilon}

\newcommand{\mathsym}[1]{{}}

\newcommand{\fin}{\begin{flushright}$\Box$\end{flushright}}
\begin{document}


\title[Spectral functions of non essentially selfadjoint operators]{Spectral functions of non essentially selfadjoint operators}

\author{H.A.\ Falomir and P.A.G.\ Pisani}

\address{ IFLP, CONICET - Departamento de F\'{\i}sica, Fac.\ de Ciencias Exactas de la UNLP, C.C. 67, (1900) La PLata, Argentina.}%

\email{falomir@fisica.unlp.edu.ar; pisani@fisica.unlp.edu.ar}%

\maketitle

\begin{abstract} One of the many problems to which J.S.\ Dowker devoted his attention is the effect of a conical singularity in the base  manifold on the behavior of the quantum fields. In particular, he studied the small-$t$ asymptotic expansion of the heat-kernel trace on a cone and its effects on physical quantities, as the Casimir energy. In this article we review some peculiar results found in the last decade, regarding the appearance of non-standard powers of $t$, and even  negative integer powers of $\log{t}$,  in this asymptotic expansion for the selfadjoint extensions of some symmetric operators with singular coefficients. Similarly, we show that the $\zeta$-function associated to these selfadjoint extensions presents an unusual analytic structure.
\end{abstract}

\date{\today}

\section{Introduction}

In Quantum Field Theory, the transition amplitude for particles interacting with a background field or subject to boundary conditions can be described in terms of the effective action of the model under study. In the one-loop approximation, this effective action can be expressed in terms of the functional determinant of the differential operator appearing in the quadratic in the quantum field term of the action. Since the spectra of these operators are unbounded, these determinants must be defined in the framework of an appropriate regularization.


In 1976, J.S.\ Dowker and R.\ Critchley \cite{Dowker} presented a powerful and elegant regularization scheme for the definition of these functional determinants, based on which is generically known as the  $\zeta$-function associated with the differential operator, built from the spectrum of the quantum fluctuations.

Since then, this formalism has been successfully applied to the determination of one-loop effective actions, vacuum energies, anomalies and other physical quantities of interest in Quantum Field Theory. It is presently a fundamental tool in the study of quantum effects in systems under the influence of external conditions \cite{Elizalde,Elizalde:1996zk,Bytsenko,Klaus,Bordag,Fursaev:2011zz}.

To be more specific, the first quantum corrections to the effective action are, in general, given by (a power which depends on the nature of the fields of) the functional determinant of an elliptic differential operator on an appropriate Hilbert space --the quantum fluctuations operator-- that is obtained as the second functional derivative of the classical action at vanishing quantum fields. J.S.\ Dowker and R.\ Critchley gave a definition of this functional determinant in terms of the derivative of the associated $\zeta$-function, thus relating the one-loop effective action with the spectrum of the quantum fluctuations operator and, consequently, with the external conditions applied on them.

This formalism is also related to the so-called proper time regularization through the relation between the $\zeta$-function and the trace of the corresponding heat-kernel \cite{Vass1} (See Appendix \ref{spectralfunctionsrelations}). Let $H$ represent an elliptic boundary problem on an $m$-di\-men\-sion\-al compact manifold $M$ with smooth boundary $\partial M$, with a discrete spectrum $\{\lambda_n\}_{n\in\mathbb{N}}$. Then, under appropriate conditions \cite{Seeley}, the $\zeta$-function is defined as the trace of the complex power $H^{-s}$,
\begin{equation}\label{zetadef}
    \zeta_H(s):={\rm Tr}\,H^{-s}:=\sum_{n\in\mathbb{N}} \lambda_n^{-s}\,,
\end{equation}
series that converges to an analytic function in an open half-plane with $\Re\left(s\right)$ large enough and admits a meromorphic extension to the whole $s$-plane. On the other hand, for positive definite $H$, the trace of the operator $e^{-tH}$ is defined as \cite{Gilkey}
\begin{equation}\label{hkdef}
    {\rm Tr}\,e^{-tH}:=\sum_{n\in\mathbb{N}} e^{-t\,\lambda_n}
\end{equation}
for $t>0$. It has been established \cite{Seeley} that for a differential operator $H$ of  order $d$ with smooth coefficients, the trace of the heat-kernel, ${\rm Tr}\,e^{-t H}$,  admits
a small-$t$ asymptotic expansion given by
\begin{equation}\label{heat-trace}
  {\rm Tr}\,e^{-t H}\sim \sum_{n=0}^\infty a_n\cdot t^{(n-m)/d}\,,
\end{equation}
where the Seeley-De Witt (SDW) coefficients $a_n$ are integrals on the manifold $M$ and
its boundary $\partial M$ of local invariants which depend on the coefficients in $H$, the metric on $M$ and the boundary conditions imposed at $\partial M$ \cite{Gilkey}. Let us remark that for a  differential operator with smooth coefficients on a base manifold $M$ without boundary the SDW coefficients $a_n$ vanish for odd $n$.

Since the function $\zeta_H(s)$ in Eq.\ (\ref{zetadef}) is related to the heat-trace ${\rm Tr}\,e^{-t H}$ in Eq.\ (\ref{hkdef}) by a Mellin transformation \cite{Vass1}, the expansion in Eq.\  (\ref{heat-trace}) implies that $\zeta_H(s)$ has isolated simple poles at
\begin{equation}
    s_n=\frac{m-n}{d}\,,\quad {\rm with}\ n=0,1,2,\ldots
\end{equation}
with residues related to the SDW coefficients by (See Appendix \ref{spectralfunctionsrelations})
\begin{equation}\label{coef-res}
{   \left.{\rm Res}
  \left[\Gamma(s)\zeta_H(s)\right]\right|_{s=s_n}= {a_n}\,.
  }
\end{equation}
In particular, $\left.{\rm Res}\left[\zeta_H(s)\right]\right|_{s=s_n}=0$ when $s_n=0,-1,-2,\dots$ and $\zeta_H(s)$ is analytic in a neighborhood of the origin.
\smallskip

The functional determinant of $H$ introduced in \cite{Dowker} can be  defined by
\begin{equation}\label{alazeta}
    \log {\rm Det} H := - \left.\frac{d}{ds} \zeta_H(s) \right|_{s\rightarrow 0}\,,
\end{equation}
where $|_{s\rightarrow 0}$ stands for the limit of the analytic continuation to a neighborhood of $s=0$.

\smallskip

When employed for the description of the one-loop contribution to the effective action, this scheme of regularization leads to local counterterms expressed in terms of the SDW coefficients (or, equivalently, in terms of  the $\zeta$-function). This justifies the intense research devoted to the $\zeta$-function and heat-kernel methods during the last decades in relation with its applications to Quantum Field Theory.

\medskip

In this context, one of the problems to which J.S.\ Dowker devoted his attention is the influence that a conical singularity in the manifold has on the quantum field behavior \cite{Dowker1,Dow,Dowker2,Dowker3,Dowker4,Dowker5,Dowker7,Dowker9}. Particles in a conic singularity, situation which can be related to the so-called Calogero models \cite{Calogero}, appear in the study of quantum fields in a black-hole background \cite{Larsen:1995ax}, in the presence of cosmic strings \cite{Fursaev:1993qk} and in condensed matter, for example. The heat-kernel trace on a conic manifold has been studied in detail \cite{checone,brusee} and shown to admit, as in the regular case, the asymptotic expansion described by Eq.\ (\ref{heat-trace}) plus a possible logarithm. However, contrary to the case of a regular background, the SDW coefficients $a_n$ do not vanish in general for odd $n$, even in the presence of no (other) boundary.

In any case, the heat-trace on a cone satisfies the asymptotic expansion in Eq.\ (\ref{heat-trace}) only if one assumes that the fields are regular at the singular point (see other studies and applications of this topic in \cite{Cognola:1993qg,Fursaev:1993qk2,Fursaev:1995ef,Fursaev:1996uz,DeNardo:1996kp,Fucci:2009hz}). In fact, if one considers other (square integrable) behaviors of the fields at the singularity one finds that these models present different spectral properties for which the expansion given in Eq.\ (\ref{heat-trace}) no longer holds.

In this article we will review some results regarding the peculiar properties of the spectral functions associated to some differential operators with regular singularities in its coefficients. In particular we will focuss on some progress made in the last years in the understanding of the heat-kernel trace and $\zeta$-function properties of some (symmetric but not essentially self adjoint) differential operators which are \emph{locally homogeneous} near the singularity, \emph{i.e.} with the singular coefficient in the \emph{potential} with the same scaling dimension as the highest derivative in the \emph{kinetic} term.
As we will see, an essential aspect of these models is the possibility of  imposing (for certain range of the parameters in the  potential) \emph{boundary conditions} at the singularity that break this scale homogeneity.

The Laplacian on a manifold with a conical singularity is an example of these kind of differential operators. The asymptotic expansion of the heat-kernel trace corresponding to this operator has been considered, probably for the first time, by A.\ Sommerfeld \cite{somcone} and H.S.\ Carlslaw \cite{carcone}. It was only in 1980 that it was pointed out by C.J.\ Callias and C.H.\ Taubes \cite{Callias1p} that, for these kind of differential operators, the heat-kernel trace small-$t$ asymptotic expansion in terms of powers of the form $t^{(n-m)/d}$ could be ill-defined and conjectured that more general powers of $t$, as well as $\log t$ terms, could appear.
Indeed, it was proved in the following years \cite{Callias1}, for some second order elliptic (essentially) selfadjoint differential operators $H$, the presence of $t^{(n-m)/2}\, \log t$ terms in the small-$t$ asymptotic expansion of the heat-kernel at the diagonal\footnote{By ``heat-kernel'' we mean the kernel $e^{-t\,H}(x,y)$ of the integral operator $e^{-t\,H}$.}, $e^{-t H}(x,x)$, with some distributional coefficients with support concentrated at the singularity.

More recently, E.\ Mooers \cite{Mooers} studied the selfadjoint extensions of the Laplacian acting on differential forms on a manifold with a conical singularity and found that the asymptotic expansion of the heat-trace contains powers of $t$ whose exponents depend on the deficiency angle of the singularity.

In similar settings, it was shown in a series of articles \cite{FPW,FMPS,Falomir:2004zf,Falomir:2005yw,Falomir:2005xh} that, for some non essentially selfadjoint locally homogeneous differential operators, the small-$t$ asymptotic expansion of the heat-kernel trace presents \emph{non-standard} powers of $t$, \emph{i.e.}  powers with exponents which are not determined by the order of the differential operator and the dimension of the base manifold only, as for the regular case (see Eq.\ (\ref{heat-trace})), but also depend on the coefficient of the singular term in the potential. Consequently, in these models, the presence of a regular singularity in the potential term of the differential operator leads to non-standard poles in the associated $\zeta$-function, which lie at positions that depend on parameters other than the order of the operator and the dimension of the manifold.

Later, K.\ Kirsten \emph{et al}.\ \cite{Kirsten:2005bh,Kirsten:2005yw,Kirsten:2007ur,Kirsten:2008wu,KL2012} considered a limit case of a symmetric second order locally homogeneous differential operator, finding that integer powers of $\log{t}$ terms appear in the small-$t$ asymptotic expansion of the heat-kernel trace and a logarithmic cut in the corresponding $\zeta$-function, so correcting an error in  Appendix A in \cite{Falomir:2004zf}.

\medskip

It is our aim to review these results concerning the non-standard behavior of the small-$t$ asymptotic expansion of the heat-kernel trace and the singularity structure of the $\zeta$-function corresponding to this kind of locally homogeneous  differential operators. We will consider the case of some symmetric operators with a regular singularity in the potential term with the same scaling dimension as the derivative term, which makes them to admit a continuous family of selfadjoint extensions (SAE) \cite{Reed-Simon}. As we will see, the (local) scaling homogeneity is in general broken in the domains of definition  of these SAE, and this fact has consequences on the behavior of the associated spectral functions which will be studied in the following.

\smallskip

In Section \ref{the-operator} we will consider a Dirac operator $D$ (see Eq.\ (\ref{D})) defined on a space of two-component functions $\Phi(x)$ with $x$ taking values in the compact segment $[0,1]$. We will introduce in the operator $D$ a singular term proportional to $1/x$, which has the same scaling dimension as the kinetic term $\partial_x$. As we will see, this simple model shows the above mentioned characteristic.

This first order differential operator is not positive definite and one can not define the associated heat-kernel. Instead, we will consider the $\zeta$-function as defined in (\ref{zetadef}) and show that it presents a non-standard pole structure due to the presence of the singular term $\sim1/x$ in $D$. In so doing, we will first determine the large-$|\lambda|$ asymptotic expansion of the resolvent $(D-\lambda)^{-1}$, since the powers of $\lambda$ in this expansion determine the position of the poles of the $\zeta$-function (see \cite{Vass1}, for example).

We will proceed as follows: Firstly, we will construct the resolvents for two particular selfadjoint extensions
for which the boundary conditions at the singular point $x=0$ are
invariant under the scaling $x \rightarrow c\,x$. The resolvent
expansion for these particular SAE displays the standard powers of $\lambda$,
leading to the standard poles for the $\zeta$-function. Secondly, we will show that the
resolvent for a general SAE is a convex linear
combination of these special resolvents with coefficients which depend on
$\lambda$. This additional dependence on $\lambda$ leads to the non-standard powers in the
resolvent asymptotic expansion of a general SAE and, hence, to non-standard poles for the
associated $\zeta$-function. The selfadjoint extensions of $D$ are not, in general, locally scale invariant at the singularity in the sense that the conditions the functions in its domain satisfy near the origin are not invariant under the scaling $x\rightarrow c\,x$. As $c \rightarrow 0$ they
tend to the conditions satisfied by the functions in the domain of one of the locally scale invariant SAE, and
as $c \rightarrow \infty$ they tend to the other. The dependence of the residues at the anomalous poles on the SAE will also be
explained by a scaling argument.

\smallskip

This model describes the central idea of our work. We consider differential operators $H$ which contain singular terms of the same scaling dimension as the highest derivative term; these operators -which we call \emph{locally homogeneous} near the singularity- admit selfadjoint extensions whose domains are characterized by \emph{boundary conditions} at the singularity that break, in general, this local homogeneity. This introduces in the definition of the SAE dimensionful parameters which are not present in its expression as a differential operator. This determines that the large-$|\lambda|$ asymptotic expansion of the resolvent-trace of a general SAE, ${\rm Tr}\,(H-\lambda)^{-1}$, presents non-standard powers of $\lambda$, with dependence on the coupling in the singular term. This, in turn, is the origin of non-standard poles of the $\zeta$-function, ${\rm Tr}\,H^{-s}$, in the complex $s$-plane and (for $H$ positive definite) non-standard powers of $t$ in the small-$t$ asymptotic expansion of the heat-kernel trace, ${\rm Tr}\,e^{-tH}$. Here, by ``non-standard'' we mean ``not determined by the dimension of the base manifold and the order of the differential operator only'', as happens for the regular case.

Indeed, in Section \ref{second-order} we consider a second order differential operator on the segment $[0,1]$ with a singular potential term of the form to $g(g-1)/x^2$. We show that, for $|g|<1/2$, one obtains similar results for the associated $\zeta$-function as those described for the first order operator case we consider in Section \ref{the-operator}. One finds that there are two locally scale invariant SAE for which the $\zeta$-function presents the usual poles. The resolvent of the general SAE is obtained as a convex linear combination of the resolvents of these particular SAE, with a coefficient dependent on the $\lambda$-parameter. This implies the presence of $g$-dependent poles in the $\zeta$-function with residues which also depend  on the SAE. On the other hand, the small-$t$ asymptotic expansion of the heat-kernel trace of the general SAE presents non-standard $g$-dependent powers of $t$ with SDW coefficients with dependence on the SAE.

This second order differential operator taken for $g=1/2$ also admits a continuous family of SAE. This is a limit case in the sense that beyond this range the operator becomes unbounded below. It has the peculiarity that only one SAE is locally scale invariant near the singularity at $x=0$. This is reflected in even more pathological properties of its associated spectral functions. In Section \ref{limitcase} we briefly review the singularities of the $\zeta$-function, which also presents a branch cut, and the behavior of the heat-kernel trace, which in the general case admits a small-$t$ asymptotic expansion in terms of negative integer powers of $\log t$.

In Section \ref{adjoint-H} we consider a locally homogeneous second order differential operator on the half-line $\mathbb{R}^+$ \cite{FPW}. In order to get discrete spectra for the SAE of this  operator we also include a quadratic term in the potential\footnote{Selfadjoint extensions for more general singular potentials have been studied in \cite{Rellich,Albeverio}.}, $V(x)=x^2+\left( \nu^2-\frac{1}{4} \right)/x^{2}$. We obtain similar results as in the compact case discussed in Section \ref{second-order}: Only two of the continuous family of SAE admitted by this Schr\"{o}dinger operator for certain range of the coupling \cite{Simon-73} have domains which remain invariant under scaling transformations. The associated spectral functions show the same properties as for the regular potential case. On the contrary, the conditions satisfied near the singularity by the functions belonging to the domains of definition  of the remaining SAE introduce a dimensionful parameter which explicitly break this local scale homogeneity, which implies the unusual behavior of the $\zeta$-function poles and powers of $t$ in the asymptotic expansion of the heat-kernel trace already described. These results are obtained employing the von Neumann's theory of selfadjoint extensions of symmetric operators \cite{Reed-Simon} to characterize the spectrum of the general SAE. They are also confirmed by an argument based on the asymptotic growth of the eigenvalues.

Let us mention that this model corresponds to a classically integrable system \cite{Perelomov} whose quantum spectrum -subject to Dirichlet boundary conditions at the singularity- has been determined in
\cite{Calogero}. This operator also appears as an effective radial Hamiltonian for an isotropic harmonic oscillator in multi-dimensional Euclidean space with given angular momentum. For this model several results are
known: the heat-kernel and the resolvent have been determined (also imposing Dirichlet boundary conditions) in \cite{Peak,Kleinert,Mueller}; the resolvent for a different boundary condition can be obtained from the Dirichlet-case by means of the so called Krein's formula \cite{Achiezer,Albeverio2}, which  will be examined in detail in Section \ref{kf}.

In Section \ref{kf} we derive a generalization of Krein's formula \cite{Krein1} which can be applied to the kind of locally homogeneous operators with singular coefficients considered in this review. Krein's formula establishes a relation between the resolvents corresponding to different selfadjoint extensions. We generalize this relation to the case of a Schr\"odinger operator with a potential that contains terms with the same  scaling  dimension as the derivative term. Our generalized Krein's formula will make manifest the presence of non-standard powers of $\lambda$ in the large-$|\lambda|$ expansion of these resolvents.


The operators considered in the present paper can be understood as the radial problem resulting from a separation of variables for particular models, and some results described in the following Sections rely on the knowledge of the behavior of the special functions appearing in their resolutions. It should be mentioned that a more general (and abstract) approach to these kind of problems has been developed in \cite{Gil1} (See also \cite{Gil2,Gil2p,Gil3,Krainer,Gil4,Gil5} and references therein). In \cite{Gil1} the asymptotic behavior of the trace of the resolvent of elliptic cone operators on compact manifolds has been considered under general conditions, showing that it admits an expansion which may show non-integer powers of $\lambda$ with coefficients which are in general rational functions of powers of $\lambda$ and $\log \lambda$, and not mere polynomials. As pointed out in this reference, this result implies that the associated $\zeta$-functions might have poles at unusual locations, or that they might even not extend meromorphically to $\mathbb{C}$ at all, which is consistent with what is described in the following.


\section{The first-order operator and its selfadjoint extensions}\label{the-operator}

Following \cite{FMPS}, in this section we consider first-order symmetric differential operator
\begin{equation}\label{D}
  D=\left(\begin{array}{cc}
    0  & \tilde{A} \\
    A & 0 \
  \end{array}\right)\, ,
\end{equation}
where
\begin{equation}\label{AA}
  A=-\partial_x + \frac{g}{x}\,,
  \qquad \tilde{A} = \partial_x + \frac{g}{x}
\end{equation}
and $|g|<1/2$, defined on the domain $\mathcal{D}(D):= \mathbb{C}^2\times\mathcal{C}_0^\infty(0,1)$, i.e.\ the set of smooth two component functions with compact support in the open segment $(0,1)$. For the given range of the coupling $g$, $D$ is not essentially selfadjoint, admitting a continuous family of SAE.

We will show that, for a general SAE of $D$, the large-$|\lambda|$ asymptotic expansion of the resolvent-trace ${\rm Tr}\{(D-\lambda)^{-1}\}$ presents non-standard powers of $\lambda$ whose exponents depend on the parameter $g$. Besides, the corresponding $\zeta$- and $\eta$-functions present simple poles lying at $g$-dependent points in the complex plane, with residues which depend on the considered SAE.

We proceed as follows: firstly, we construct the resolvents for two particular SAE for which the behavior of the functions in the definition domain of the SAE near the singular point $x=0$ is invariant under the scaling $x \rightarrow c\,x$. The asymptotic expansion of the resolvent for these special extensions displays the usual powers of $\lambda$, then leading to the usual poles for the corresponding $\zeta$- and $\eta$-functions.  Secondly, we show that the resolvents for the remaining SAE are convex linear combinations of these special extensions with $\lambda$-dependent coefficients. This additional dependence on $\lambda$ leads to non-standard powers in the asymptotic expansion of the resolvent, and hence to non-standard poles for the $\zeta$- and $\eta$-functions.

These SAE are not locally invariant under the scaling $x\rightarrow c\,x$ in the sense that the conditions defining the behavior of the functions belonging to its domain near $x=0$ are not scale-invariant. Rather, as $c \rightarrow 0$ they
tend to the conditions for the domain of one of the locally scale invariant SAE, while as $c \rightarrow \infty$ they tend to the other. As a consequence, the residues at the anomalous poles of the $\zeta$-function tend to zero as $c
\rightarrow 0$ and diverge as $c \rightarrow \infty$. This behavior will be explained by means of a scaling argument.

\smallskip

Integration by parts shows that the operator $D$ is symmetric on $\mathcal{D}(D)$. The adjoint operator $D^\dagger$, which is the maximal extension of $D$, is defined on the domain $\mathcal{D}(D^\dagger)$ of functions
$\Phi(x)= \left(\begin{array}{c}
  \phi_1(x) \\
  \phi_2(x)
\end{array}\right)\in \mathbb{C}^2\times\mathbf{L_2}(0,1)$ having a locally sumable
first derivative and such that
\begin{equation}\label{DPhi}
    D\Phi(x)=\left( \begin{array}{c}
      \tilde{A} \phi_2(x) \\
      A \phi_1(x)
    \end{array} \right) = \left(\begin{array}{c}
      f_1(x) \\
      f_2(x)
    \end{array}\right)\in \mathbb{C}^2\times\mathbf{L_2}(0,1)\, .
\end{equation}
Since $D$ is symmetric, $\mathcal{D}(D)\subset\mathcal{D}(D^\dagger)$. The following lemma characterizes the behavior at the singular point $x=0$ of the two components of the functions in $\mathcal{D}(D^\dagger)$.

\begin{lem} \label{lema1-1}
If $\Phi(x)\in \mathcal{D}(D^\dagger)$ and $-\frac{1}{2}< g<
\frac{1}{2}$, then
\begin{equation}\label{lemaI}
  \big|\,\phi_1(x)-C_1[\Phi]\, x^g\big| +
  \big|\,\phi_2(x)-C_2[\Phi]\, x^{-g}\big|
  \leq K_g\, \|D\Phi(x)\| \, x^{1/2}\, ,
\end{equation}
for some constants $K_g$, $C_1[\Phi]$ and $C_2[\Phi]$, where $\| \cdot
\|$ is the $\mathbf{L_2}$-norm.
\end{lem}

\noindent{\bf Proof:} Eq.\ (\ref{DPhi}) implies
\begin{eqnarray}\label{phi-chi-en0}
      \phi_1(x)= C_1[\Phi]\, x^g - x^g \, \int_0^x y^{-g}\, f_2(y)\, dy
     \,  , \\
     \phi_2(x)= C_2[\Phi]\, x^{-g} + x^{-g} \, \int_0^x y^{g}\, f_1(y)\, dy
     \, ,
\end{eqnarray}
and taking into account that
\begin{eqnarray}\label{schwarz}
      \left|\int_0^x y^{g} \, f_1(y)\, dy\right|\leq
      \frac{x^{g+1/2}}{\sqrt{1+2g}}\, \|f_1\| \, ,
\\
      \left|\int_0^x y^{-g} \, f_2(y)\, dy\right|\leq
      \frac{x^{-g+1/2}}{\sqrt{1-2g}}\, \|f_2\| \,  ,
\end{eqnarray}
we immediately get Eq.\ (\ref{lemaI}) with $K_g =
(1-2g)^{-1/2}+(1+2g)^{-1/2}$.\fin

\bigskip

The following lemma will be useful to describe the selfadjoint extensions of $D$.
\begin{lem}
Let $\Phi(x)=\left(\begin{array}{c}
  \phi_1(x) \\
  \phi_2(x)
\end{array}\right),\Psi(x)=\left(\begin{array}{c}
  \psi_1(x) \\
  \psi_2(x)
\end{array}\right) \in \mathcal{D}(D^\dagger)$. Then
\begin{eqnarray}\label{DDstar}
    \left(D^\dagger \Psi, \Phi\right) -
  \left(\Psi, D^\dagger \Phi\right) = \\\nonumber
  = \Big\{
  C_1[\Psi]^* C_2[\Phi] - C_2[\Psi]^* C_1[\Phi]\Big\}+
  \Big\{\psi_2(1)^*\, \phi_1(1)
  - \psi_1(1)^*\, \phi_2(1) \Big\}\, .
\end{eqnarray}
\end{lem}

\noindent{\bf Proof:} From eqs.\ (\ref{AA}) one easily obtains
\begin{eqnarray}\label{DDstar2}
      \left(D ^\dagger\Psi, \Phi\right) -
  \left(\Psi, D^\dagger \Phi\right) = \\
    \lim_{\varepsilon \rightarrow 0^+}\int_\varepsilon^1
    \partial_x \Big\{x^g\,\psi_2(x)^*\, x^{-g}\,\phi_1(x)
    - x^{-g}\,\psi_1(x)^* \, x^g \, \phi_2(x)
    \Big\} dx\, ,
\end{eqnarray}
from which, taking into account the results in Lemma
\ref{lema1-1}, Eq.\ (\ref{DDstar}) follows directly.\fin

\bigskip

Next, if $\Psi(x)$ in Eq.\ (\ref{DDstar}) belongs to the domain of
the closure of $D$, $\overline{D}=(D^{\dagger})^\dagger$, i.\ e.\
\begin{equation}\label{en-la-clausura}
  \Psi(x)\in
\mathcal{D}(\overline{D}) \subset \mathcal{D}(D^\dagger)\,  ,
\end{equation}
then the right hand side of Eq.\ (\ref{DDstar}) must vanish for
any $\Phi(x)\in \mathcal{D}(D^\dagger)$. Therefore
\begin{equation}\label{Psi-clausura}
    C_1[\Psi]=C_2[\Psi]=\Psi(1)=0\, .
\end{equation}
On the other hand, if $\Psi(x)$ and $\Phi(x)$ belong to the domain of a
symmetric extension of $D$ -which must be contained in $\mathcal{D}(D^\dagger)$- then
the right hand side of Eq.\ (\ref{DDstar}) must also vanish. Thus, each closed extension of $D$ corresponds to a subspace
of $\mathbb{C}^4$ under the map $\Phi\rightarrow \left( C_1[\Phi],
C_2[\Phi], \phi_1(1), \phi_2(1) \right)$. If we define the orthogonal complement in
terms of the symplectic  form of the right hand side of Eq.\
(\ref{DDstar}), then the selfadjoint
extensions correspond to those subspaces $S\subset \mathbb{C}^4$
such that $S=S^{\perp}$.

Since we are interested in the consequences of the singularity at the origin, for simplicity in the following we will consider SAE satisfying the local boundary condition
\begin{equation}\label{BC1}
  \phi_1(1)=0\, .
\end{equation}
Each extension is then determined by a condition of the form
\begin{equation}\label{BC2}
    \alpha\, C_1[\Phi] + \beta\, C_2[\Phi] = 0\, ,
\end{equation}
with $\alpha,\beta \in \mathbb{R}$ and $\alpha^2 + \beta^2 =1
$. We denote this selfadjoint extension by $D^{(\alpha, \beta)}$.

\subsection{The spectrum} \label{the-spectrum}

In order to determine the spectrum of an arbitrary selfadjoint extension $D^{(\alpha,\beta)}$, we need the formal solutions of
\begin{equation}\label{Ec-hom}
  (D-\lambda)\Phi(x)=0 \Rightarrow \left\{
  \begin{array}{c}
    \tilde{A} \phi_2(x) = \lambda \phi_1(x)\, ,\\ \\
    A \phi_1(x) = \lambda \phi_2(x)\, ,
  \end{array}\right.
\end{equation}
satisfying the boundary conditions given by eqs.\ (\ref{BC1}) and
(\ref{BC2}).

The zero-mode, corresponding to $\lambda=0$, is given by
\begin{equation}\label{lambda0}
  \Phi(x)=
\begin{pmatrix}
  C_1 \, x^g\\
  C_2 \, x^{-g}
\end{pmatrix}\, .
\end{equation}
The boundary condition at $x=1$ chosen by Eq.\ (\ref{BC1}) implies $C_1=0$. The boundary condition at $x=0$ given by Eq.\  (\ref{BC2}) tells that $C_2=0$ unless  $\beta = 0$. Consequently, only the selfadjoint extension $D_x^{(1,0)}$ has a zero-mode.

For $\lambda\neq 0$ we apply $\tilde{A}$ to the second line in Eq.\ (\ref{Ec-hom}) so that, using the first line in Eq.\ (\ref{Ec-hom}), one gets
\begin{equation}\label{hom-sec}
    \left\{ \partial_x^2 - \frac{g(g-1)}{x^2}
    + \lambda^2\right\} \phi_1(x)=0\, .
\end{equation}
The solutions of this equation take the form
\begin{equation}\label{sol-hom-1}
  \phi_1(x) = K_1\, \sqrt{X}\, J_{\frac 1 2 -g}(X)
  + K_2 \, \sqrt{X} \, J_{g-\frac 1 2}(X)\, ,
\end{equation}
where we have made the rescaling $X:= \tilde{\lambda} \, x$ and defined
$\tilde{\lambda}=+\sqrt{\lambda^2}$; $K_1, K_2$ are complex constants. The lower component results
\begin{equation}\label{hom-2}
  \phi_2(x)= \sigma
  \left\{-K_1 \, \sqrt{X}\, J_{-g-\frac 1 2}(X)
  + K_2\, \sqrt{X} \, J_{g+\frac 1 2}(X)\right\}\, ,
\end{equation}
where $\sigma={\tilde{\lambda}}/{\lambda}$.

From the behavior of $\phi_1(x),\phi_2(x)$ at the origin we can get $C_1[\Phi],C_2[\Phi]$, respectively. The boundary condition now reads
\begin{equation}\label{ec-spectrum}
    \alpha\, C_1[\Phi] + \beta\, C_2[\Phi]
    =
    \frac{\alpha\, K_2\, \tilde{\lambda}^g}
    {2^{g-\frac{1}{2}}
  \Gamma\left(\frac{1}{2}+g\right)}
  - \sigma
  \frac{\beta\, K_1\,
  \tilde{\lambda}^{-g}}{2^{-g-\frac{1}{2}}
  \Gamma\left(\frac{1}{2}-g\right)}=0\, .
\end{equation}

Let us first consider a particular selfadjoint extension, namely $\alpha=0$ and $\beta=1$. For $\alpha = 0$, Eq.\ (\ref{ec-spectrum}) implies $K_1 = 0$.
Therefore, from the condition $\phi_1(1)=0$ we obtain the spectrum equation $J_{g-\frac 1
2}(\tilde{\lambda})=0$. Thus the spectrum of $D_x^{(0,1)}$ is given by
\begin{equation}\label{alpha0}
  \lambda_{\pm,n} = \pm j^{(n)}_{g-\frac 1 2}\, ,
   \quad n=1,2,\dots
\end{equation}
where $j^{(n)}_{\nu}$ is the $n$-th positive zero of the Bessel
function $J_{\nu}(z)$. The spectrum is non-degenerate and symmetric with respect to the origin.

For $\alpha\neq 0$, Eq.\ (\ref{ec-spectrum}) can be written as
\begin{equation}\label{spectrum}
  \frac{K_2}{K_1} = \sigma\,
  \tilde{\lambda}^{-2 g} \left[
  \frac{4^{g}\, \Gamma\left(\frac{1}{2}+g\right)}
  {\Gamma\left(\frac{1}{2}-g\right)}\right]
  \left(\frac{\beta}{\alpha}\right)\, .
\end{equation}
In this case, the boundary condition at $x=1$ determines the
eigenvalues as the solutions of the transcendental equation
\begin{equation}\label{eigenvalues}
  \tilde{\lambda}^{2 g}\,
  \frac{J_{\frac 1 2 -g}(\tilde{\lambda})}
  {J_{g-\frac 1 2}(\tilde{\lambda})}
  =\sigma \, \rho(\alpha,\beta)\, ,
\end{equation}
where we have defined
\begin{equation}\label{rho}
  \rho(\alpha,\beta):=
   -\frac{4^{g}\, \Gamma\left(\frac{1}{2}+g\right)}
  {\Gamma\left(\frac{1}{2}-g\right)}\,
  \left(\frac{\beta}{\alpha}\right)\, .
\end{equation}

\begin{figure}
    \epsffile{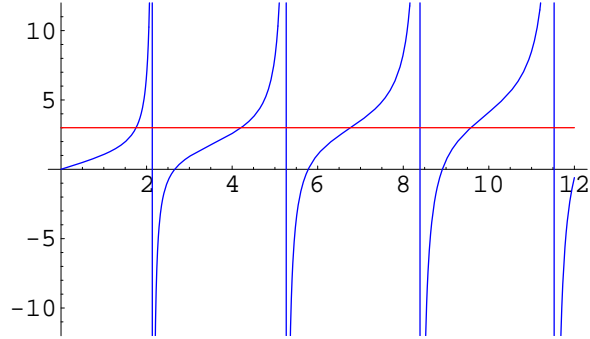}
    \caption{Plot for $F(\lambda):= \displaystyle{\lambda^{2 g}\,
  \frac{J_{\frac 1 2 -g}(\lambda)}{J_{g-\frac 1 2}(\lambda)}
  }$ for $g=1/3$ and $\rho(\alpha,\beta)=3$.} \label{figure}
\end{figure}

For the positive eigenvalues $\tilde{\lambda}=\lambda$, $\sigma =1 $ and Eq.\ (\ref{eigenvalues}) reduces to
\begin{equation}\label{eigenvalues-pos}
  F(\lambda):=\lambda^{2 g}\,
  \frac{J_{\frac 1 2 -g}(\lambda)}{J_{g-\frac 1 2}(\lambda)}
  = \rho(\alpha,\beta)\,.
\end{equation}
This expression has been plotted in Figure \ref{figure} for a particular value of
$\rho(\alpha,\beta)$ and $g$.

On the other hand, for negative eigenvalues $\lambda = e^{i\, \pi}
\tilde{\lambda}$, $\sigma =e^{-i\, \pi}$ and Eq.\ (\ref{eigenvalues}) reads
\begin{equation}\label{eigenvalues-neg}
  F(\tilde{\lambda})= e^{-i\, \pi} \rho(\alpha,\beta)
  =\rho(\alpha,-\beta)\, .
\end{equation}
Therefore, the negative eigenvalues of $D^{(\alpha, \beta)}$ have the same absolute value as the positive eigenvalues of $D^{(\alpha,
-\beta)}$.

Notice that for any selfadjoint extension the spectrum is non-degenerate and that there is a
positive eigenvalue between each pair of consecutive zeroes of
$J_{g-\frac 1 2}(\lambda)$. Moreover, the spectrum is symmetric with respect to the origin
only for the selfadjoint extension corresponding to $\alpha=0$ (from now on, the
``D-extension", see Eq.\ (\ref{alpha0})) and for the extension corresponding to $\beta=0$ (which we will call the ``N-extension"). Indeed, from eqs.\ (\ref{eigenvalues}) and (\ref{rho}) one can see
that the eigenvalues of $D_x^{(1,0)}$ are given by
\begin{equation}\label{eigen-beta0}
  \lambda_0 =0\, , \quad
  \lambda_{\pm,n} = \pm j^{(n)}_{\frac 1 2 -g}\, ,\quad \ n=1,2,\dots
\end{equation}

\subsection{The resolvent} \label{the-resolvent}

In this section we will construct the resolvent of $D$
\begin{equation}\label{def-resolv}
  G(\lambda)=(D-\lambda)^{-1}\, ,
\end{equation}
for its different selfadjoint extensions.

We will first consider the two limiting cases in Eq.\ (\ref{BC2}),
namely the ``$D$-extension" and the ``$N$-extension". The resolvent for a general selfadjoint extension
will be later evaluated as a linear combination of those obtained
for these two limiting cases.

The kernel of the resolvent 
\begin{equation}\label{resolv}
  G(x,y; \lambda)=\left(\begin{array}{cc}
    G_{11}(x,y; \lambda) & G_{12}(x,y; \lambda) \\
    G_{21}(x,y; \lambda) & G_{22}(x,y; \lambda)
  \end{array}\right)
\end{equation}
satisfies
\begin{equation}\label{G}
   (D-\lambda)\, G(x,y; \lambda) =
\delta(x,y)\, \mathbf{1}_2\, ,
\end{equation}
from which we straightforwardly  get for the diagonal elements
\begin{equation}\label{ec-dif-g-diag}
  \begin{array}{c}
  \displaystyle{
    \left\{ \partial_x^2 - \frac{g(g-1)}{x^2}
    + \lambda^2 \right\}  G_{11}(x,y; \lambda) = -\lambda\,
     \delta(x,y)\, ,} \\\\
       \displaystyle{
       \left\{ \partial_x^2 - \frac{g(g+1)}{x^2}
    + \lambda^2 \right\}  G_{22}(x,y; \lambda) = -\lambda\,
     \delta(x,y) }\, ,
  \end{array}
\end{equation}
while for the non diagonal ones we have
\begin{equation}\label{ec-dif-g-nodiag}
  \begin{array}{c}
      \displaystyle{
    G_{21}(x,y; \lambda)= \frac{1}{\lambda}
    \left\{- \partial_x + \frac g x \right\}
    G_{11}(x,y; \lambda)\, , }\\ \\
      \displaystyle{
    G_{12}(x,y; \lambda)= \frac{1}{\lambda}
    \left\{ \partial_x + \frac g x \right\}
    G_{22}(x,y; \lambda)\, ,}
  \end{array}
\end{equation}
assuming $\lambda\neq 0$.

Since the resolvent is analytic in $\lambda$, it is sufficient to
evaluate it on the open right half-plane. In order to do that, we use the upper and lower components of some
particular solutions of the homogeneous equation (\ref{Ec-hom}).

Let us define
\begin{equation}\label{soluciones}
    \left\{
  \begin{array}{l}
    L_1^D(X)= \sqrt{X}\, J_{g-\frac 1 2}(X)\,  ,\\ \\
        L_2^D(X)= \sqrt{X}\, J_{g+\frac 1 2}(X)\,  ,\\ \\
        L_1^N(X)= \sqrt{X}\, J_{\frac 1 2 -g}(X)\,  ,\\ \\
        L_2^N(X)= \sqrt{X}\, J_{-g-\frac 1 2}(X)\, , \\ \\
    R_1(X;\lambda)=\sqrt{X} \left[J_{g-\frac 1 2}({\lambda})
     J_{\frac 1 2 -g}(X)
    - J_{\frac 1 2 -g}({\lambda}) J_{g-\frac 1 2}(X)\right]\, ,
    \\ \\
    R_2(X;\lambda)=\sqrt{X} \left[J_{g-\frac 1 2}({\lambda})
    J_{-g-\frac 1 2}(X)
    + J_{\frac 1 2 -g}({\lambda}) J_{g+\frac 1 2}(X)\right]\, .
  \end{array}
  \right.
\end{equation}
Notice that $R_1({\lambda};\lambda)=\left.\tilde{A}\,
R_2({\lambda}\, x;\lambda )\right|_{x=1}=0$. We will also need the Wronskians
\begin{equation}\label{W-D}
\left\{
\begin{array}{c}
  \displaystyle{
  W\left[ L_1^D(X), R_1(X;\lambda) \right] =
   - \frac 2 \pi \, \cos(g\, \pi) \,
  J_{g-\frac 1 2}({\lambda})=:\frac{1}{ \gamma_D(\lambda)}\, ,
    } \\ \\
  \displaystyle{
  W\left[ L_2^D(X), R_2(X;\lambda) \right] =
   \frac 2 \pi \, \cos(g\, \pi) \,
  J_{g-\frac 1 2}({\lambda})=:\frac{-1}{ \gamma_D(\lambda)}\, ,
  } \\ \\
    \displaystyle{
  W\left[ L_1^N(X), R_1(X;\lambda) \right] =
   - \frac 2 \pi \, \cos(g\, \pi) \,
  J_{\frac 1 2 -g}({\lambda})=:\frac{1}{ \gamma_N(\lambda)}\, ,
    } \\ \\
  \displaystyle{
  W\left[ L_2^N(X), R_2(X;\lambda) \right] =
   - \frac 2 \pi \, \cos(g\, \pi) \,
  J_{\frac 1 2-g}({\lambda})=:\frac{1}{ \gamma_N(\lambda)}
  }\, ,
  \end{array}
  \right.
\end{equation}
which  vanish only at the zeroes of $J_{\nu}({\lambda})$, for $\nu
= \pm\left(\frac 1 2 -g\right)$.

\bigskip

\subsection{The resolvent for the $D$-extension} In this case,
the function
\begin{equation}\label{func-inhom}
  \Phi(x) = \int_0^1 G_D( x, y; \lambda)
  \left( \begin{array}{c}
    f_1(y) \\
    f_2(y)
  \end{array} \right) dy
\end{equation}
must satisfy the boundary conditions $\phi_1(1)=0$ and $C_2[\Phi]=0$, for any functions
$f_1(x),f_2(x)\in \mathbf{L_2}(0,1)$.

This requires that
\begin{equation}\label{GD11}
  G_{11}^D (x,y; \lambda)= \gamma_D(\lambda) \times  \left\{
  \begin{array}{c}
    L_1^D(X)\, R_1(Y; \lambda),\ {\rm for}\ x\leq y\, , \\ \\
    R_1(X; \lambda)\, L_1^D(Y), \ {\rm for}\ x \geq y\, ,
  \end{array}\right.
\end{equation}
and
\begin{equation}\label{GD22}
  G_{22}^D (x,y; \lambda)= - \gamma_D(\lambda) \times  \left\{
  \begin{array}{c}
    L_2^D(X)\, R_2(Y; \lambda),\ {\rm for}\ x\leq y\, , \\ \\
    R_2(X; \lambda)\, L_2^D(Y), \ {\rm for}\ x \geq y\, ,
  \end{array}\right.
\end{equation}
whereas the components $G_{12}^D(x,y; \lambda)$ and
$G_{21}^D(x,y; \lambda)$ are given by Eq.\
(\ref{ec-dif-g-nodiag}).

The fact that the boundary conditions are
satisfied, as well as $(D-\lambda)\, \Phi(x)=\begin{pmatrix}
  f_1(x) \\
  f_2(x)
\end{pmatrix}$, can be straightforwardly verified from eqs.\
(\ref{soluciones}) and (\ref{W-D}). Indeed, from eqs.\ (\ref{ec-dif-g-nodiag} - \ref{GD22}) one
gets
\begin{equation}\label{near0-D}
    \phi_1(x)=
    C_1^D[\Phi] \, x^g + O(\sqrt{x})\, ,\quad
    \phi_2(x)= O(\sqrt{x})\,  ,
\end{equation}
with
\begin{equation}\label{C+}\begin{array}{c}
  \displaystyle{C_1^D[\Phi] =   \frac{-\, \pi\, \lambda^{g+1}}
    {2^{\frac{1}{2}+ g} \cos(g\, \pi)
    J_{g-\frac{1}{2}}({\lambda}) \,
    \Gamma\left(\frac{1}{2} +g\right)}}
    \times  \\ \\
  \displaystyle{  \times
    \int_0^1 \Big[R_1(\lambda\, y; \lambda) f_1(y)
    - R_2(\lambda\, y; \lambda) f_2(y)\Big] dy}\, ,
\end{array}
\end{equation}
for $\lambda$ not a zero of $J_{g-\frac{1}{2}}({\lambda})$.

Notice that $C_1^D[\Phi]\neq 0$ if the integral in the right hand
side of Eq.\ (\ref{C+}) is non vanishing.

\subsection{The resolvent for the $N$-extension} In this case,
the function
\begin{equation}\label{func-inhom-N}
  \Phi(x) = \int_0^1 G_N(x, y; \lambda)
  \left( \begin{array}{c}
    f_1(y) \\
    f_2(y)
  \end{array} \right) dy
\end{equation}
satisfies the boundary conditions $\phi_1(1)=0$ and $C_1[\Phi]=0$, for any functions
$f_1(x),f_2(x)\in \mathbf{L_2}(0,1)$.

This requires that
\begin{equation}\label{GN11}
  G_{11}^N (x,y; \lambda)= \gamma_N(\lambda) \times  \left\{
  \begin{array}{c}
    L_1^N(X)\, R_1(Y; \lambda),\ {\rm for}\ x\leq y\, , \\ \\
    R_1(X; \lambda)\, L_1^N(Y), \ {\rm for}\ x \geq y\, ,
  \end{array}\right.
\end{equation}
and
\begin{equation}\label{GN22}
  G_{22}^N (x,y; \lambda)= \gamma_N(\lambda) \times  \left\{
  \begin{array}{c}
    L_2^N(X)\, R_2(Y; \lambda),\ {\rm for}\ x\leq y\, , \\ \\
    R_2(X; \lambda)\, L_2^N(Y), \ {\rm for}\ x \geq y\, ,
  \end{array}\right.
\end{equation}
whereas the components $G_{12}^N(x,y; \lambda)$ and
$G_{21}^N(x,y; \lambda)$ are given by Eq.\
(\ref{ec-dif-g-nodiag}).

These boundary conditions, as well as the
fact that $(D-\lambda)\, \Phi(x)=\begin{pmatrix}
  f_1(x) \\
  f_2(x)
\end{pmatrix}$, can be
straightforwardly verified from eqs.\ (\ref{soluciones}) and (\ref{W-D}). In this case, from eqs.\  (\ref{ec-dif-g-nodiag} - \ref{W-D}) and (\ref{func-inhom-N} - \ref{GN22}) one
gets
\begin{equation}\label{near0-N}
    \phi_1(x)=
    O(\sqrt{x})\, ,\quad
    \phi_2(x)= C_2^N[\Phi] \, x^{-g} + O(\sqrt{x})\, ,
\end{equation}
with
\begin{equation}\label{C-}\begin{array}{c}
  \displaystyle{C_2^N[\Phi] =   \frac{\pi\, \lambda^{1-g}}
    {2^{\frac{1}{2}- g} \cos(g\, \pi)
    J_{\frac{1}{2}-g}({\lambda}) \,
    \Gamma\left(\frac{1}{2}-g\right)} \times} \\ \\
  \displaystyle{ \phantom{C_2[\Phi] = } \times
    \int_0^1 \Big[R_1(\lambda\, y; \lambda) f_1(y)
    - R_2(\lambda\, y; \lambda) f_2(y)\Big] dy\, ,}
\end{array}
\end{equation}
for $\lambda$ not a zero of $J_{\frac{1}{2}-g}({\lambda})$.

Notice that $C_2^N[\Phi]\neq 0$ if the integral in the right hand
side of Eq.\ (\ref{C-}) (the same integral as the one appearing in
the $D$-extension, Eq.\ (\ref{C+})) is non vanishing.

\subsection{The resolvent for a general selfadjoint extension of $D^{(\alpha,\beta)}$}

For the general case, we can implement the boundary conditions
\begin{equation}\label{BC-general}
  \phi_1(1)=0\, ,\quad \alpha\, C_1[\Phi] + \beta\, C_2[\Phi] =
  0\, ,\ \alpha,\beta\neq 0
\end{equation}
on
\begin{equation}\label{func-inhom-gen}
  \Phi(x) = \int_0^1 G( x,  y; \lambda)
  \left( \begin{array}{c}
    f_1(y) \\
    f_2(y)
  \end{array} \right) dy
\end{equation}
for any $f_1(x),f_2(x)\in \mathbf{L_2}(0,1)$ by taking a linear
combination of the $D$- and $N$-resolvents
\begin{equation}\label{linear-comb}
  G(x,y; \lambda)= \left[1- \tau(\lambda)\right] G_D(x,y; \lambda) +
  \tau(\lambda)\, G_N(x,y; \lambda)\, .
\end{equation}

Since the boundary condition at $x=1$ is already satisfied,
we must now impose
\begin{equation}\label{ec-tau}
  \alpha \left[1- \tau(\lambda)\right] C_1^D[\Phi]
  + \beta\, \tau(\lambda)\, C_2^N[\Phi] =0\, .
\end{equation}

Notice that, by virtue of eqs.\ (\ref{C+}), (\ref{C-}) and
(\ref{eigenvalues-pos}),
\begin{equation}\label{nocero}
  \alpha \, C_1^D[\Phi]-\beta\,  C_2^N[\Phi]=0
\end{equation}
whenever $\lambda$ is an eigenvalue of
$D^{(\alpha,\beta)}$. Therefore, from Eq.\ (\ref{ec-tau}) we get
the resolvent of $D^{(\alpha,\beta)}$ by setting
\begin{equation}\label{taudelambda}\begin{array}{c}
    \tau(\lambda) = \displaystyle{\frac{\alpha \, C_1^D[\Phi]}
  {\alpha \, C_1^D[\Phi]-\beta\,  C_2^N[\Phi]}} =
     \frac{1}{1-\displaystyle{
        \frac{\rho(\alpha,\beta)}{F(\lambda)}}} =
  \\  \\ =
  1-\displaystyle{ \frac{1}
  {1 - \displaystyle{\frac{{\lambda }^{2\,g}}{\rho(\alpha,\beta)}\
  \frac{ J_{\frac{1}
          {2} - g}(\lambda )}{
       J_{g-\frac{1}{2}}(\lambda )} }} \, ,
          }
\end{array}
\end{equation}
for  $\lambda$ not a zero of $J_{g-\frac{1}{2}}(\lambda )$.

\subsection{The trace of the resolvent} \label{trace-resolvent}

It follows from Eq.\ (\ref{linear-comb}) that the resolvent of a
general selfadjoint extension of $D$ can be expressed in terms
of the resolvents of the two limiting cases, $G_D(\lambda)$ and
$G_N(\lambda)$. Moreover, since the eigenvalues of any extension
grow linearly with $n$ (see section \ref{the-spectrum}), these
resolvents are Hilbert-Schmidt operators and their
$\lambda$-derivatives are trace class.

From the relation
\begin{equation}\label{derivG}\begin{array}{c}
    {G(\lambda)}^2=\partial_\lambda G(\lambda) =
    \partial_\lambda G_D(\lambda) - \\ \\
  - \tau'(\lambda) \left[G_D(\lambda)-G_N(\lambda)\right]
  - \tau(\lambda)
  \left[ \partial_\lambda G_D(\lambda) -
  \partial_\lambda G_N(\lambda) \right]\, ,
\end{array}
\end{equation}
it follows that the difference
$G_D(\lambda)-G_N(\lambda)$ is a strongly analytic function of
$\lambda$ (except at the zeroes of $\tau'(\lambda)$) taking values
in the trace class operators ideal.

From the explicit expressions of $G_D(\lambda)$ and
$G_N(\lambda)$ (see Eqs.\ (\ref{GD11}),
(\ref{GD22}), (\ref{GN11}) and (\ref{GN22})) we straightforwardly
get
\begin{eqnarray}\label{traza-derivGD}
        {\rm Tr}\{\partial_\lambda G_D(\lambda)\} =
      \int_0^1
      {\rm tr}\{
      \partial_\lambda G_D( x, x; \lambda)\}\, dx =
    \nonumber\\
    =\partial_\lambda
      \left\{\frac{J_{g+ \frac 1 2 }(\lambda)}
      {J_{g-\frac 1 2}(\lambda)}\right\} =
      1 - \frac{g^2}{{\lambda }^2} +
  {\left( \frac{1}{2\,\lambda } +
      \frac{{J'_{g- \frac{1}{2} }}( \lambda )}
      {J_{g- \frac{1}{2}}(\lambda )} \right) }^2\, .
\end{eqnarray}
Similarly,
\begin{equation}\label{TrGD-GN}
      {\rm Tr}\{G_D(\lambda)-G_N(\lambda)\}
    =-\frac{2\,g}{\lambda } -
  \frac{{J'_{\frac{1}{2} - g}}(\lambda )}
  {J_{\frac{1}{2} - g}(\lambda )} +
  \frac{{J'_{g- \frac{1}{2}}}(\lambda )}
  {J_{g-\frac{1}{2}}(\lambda )}
    \, .
\end{equation}
Moreover, since
\begin{equation}\label{derTr}
  \partial_\lambda {\rm Tr}\{G_D(\lambda)-G_N(\lambda)\}
  ={\rm Tr}\{\partial_\lambda G_D(\lambda) -
    \partial_\lambda G_N(\lambda)\}\, ,
\end{equation}
we get
\begin{eqnarray}\label{trazas}
    {\rm Tr}\{\partial_\lambda G_D(\lambda) -
    \partial_\lambda G_N(\lambda)\}
    =\nonumber\\
    =    \frac{2\,g}{{\lambda }^2} +
  {\left( \frac{1}{2\,\lambda } +
      \frac{{J'_{\frac{1}{2} - g}}(
         \lambda )}{J_{\frac{1}{2} - g}(\lambda )}
      \right) }^2 - {\left( \frac{1}
       {2\,\lambda } +
      \frac{{J'_{g- \frac{1}{2}}}( \lambda )}
      {J_{ g- \frac{1}{2}}( \lambda )} \right) }^2\, .
\end{eqnarray}
Finally, we can also write
\begin{equation}\label{TrG2}
  {\rm Tr}\{{G(\lambda)}^2\} = {\rm Tr}\{\partial_\lambda G_D(\lambda)\} -
  \partial_\lambda \Big[ \tau(\lambda)\,
  {\rm Tr}\{G_D(\lambda)-G_N(\lambda)\}\Big]\, .
\end{equation}

\subsection{Asymptotic expansion for the trace of the
resolvent}\label{Asymptotic-expansion}

Using the Hankel asymptotic expansion for the Bessel functions
\cite{A-S} (see Appendix \ref{Hankel}), we get for the first term
in the right hand side of Eq.\ (\ref{TrG2})
\begin{equation}\label{asymp-trGD-upp}\begin{array}{c}
    \displaystyle{{\rm Tr}\{\partial_\lambda G_D(\lambda)\} \sim
  \sum_{k=2}^\infty \frac{A_k(g,\sigma)}{\lambda^k}
  = }\\ \\ =\displaystyle{
   - \frac{g }
   {{\lambda }^2} + i\,\sigma\,
  \frac{g\,\left(  g -1 \right)}{{\lambda }^3} -
  \frac{3}{2}\, \frac{
      g\,\left( g  -1 \right)}{{\lambda }^4} \, +}
      \\ \\ \displaystyle{  + i \, \sigma\,
  \frac{\left( g -3  \right) \,
     \left(  g -1 \right) \,g\,
     \left(  g+  2 \right) }{2\,{\lambda }^5} +
   {{O}\left(\frac{1}
      {\lambda }\right)}^6
    }\, ,
\end{array}
\end{equation}
where $\sigma = 1$ for $\Im(\lambda)>0$, and $\sigma = -1$ for
$\Im(\lambda)<0$. The coefficients in this series can be
straightforwardly evaluated from eqs.\ (\ref{traza-derivGD}) and (\ref{JprimasobreJ}).
Notice that $A_k(g,-1)=A_k(g,1)^*$, since $A_{2k}(g,1)$ is real
and $A_{2k+1}(g,1)$ is purely imaginary.

Similarly, from Eqs.\ (\ref{TrGD-GN}), (\ref{trazas}) and
(\ref{JprimasobreJasymp}) we get
\begin{eqnarray}\label{trdifasymp}
   Tr\{G_D(\lambda) -
    G_N(\lambda)\} \sim - \frac{2 g}{\lambda}\,,\\\label{trdifderasymp}
   Tr\{\partial_\lambda G_D(\lambda) -
    \partial_\lambda G_N(\lambda)\} \sim
    \frac{2 g}{\lambda^2}\, .
\end{eqnarray}

On the other hand, taking into account Eq.\ (\ref{JsobreJup}), we
have
\begin{equation}\label{tau-asymp}
    \tau(\lambda) \sim   \left\{
  \begin{array}{l}
      {\displaystyle -\sum_{k=1}^\infty \left( \frac{e^{\sigma\, i\, \pi\, (\frac 1 2 -g)}
    \lambda^{2 g}}{\rho(\alpha,\beta)}\right)^k , \ {\rm for}\
    -\frac 1 2 < g < 0\, ,} \\ \\
      \displaystyle{
    \sum_{k=0}^\infty \left(\rho(\alpha,\beta)\,
    {e^{-\sigma\, i\, \pi\, (\frac 1 2 -g)} \,
    \lambda^{-2 g}}\right)^k , \ {\rm for}\
    0 < g < \frac 1 2\, ,}
  \end{array}
  \right.
\end{equation}
where $\sigma =1$ ($\sigma =-1$) corresponds to $\Im(\lambda)>0$
($\Im(\lambda)<0$). Notice the appearance of non integer,
$g$-dependent powers of $\lambda$ in this asymptotic expansion.

Similarly
\begin{equation}\label{tauprima}
    \tau'(\lambda)
   \sim
   \left\{\begin{array}{l}
     \displaystyle{
    - \frac{2\,g}{\lambda}\,\sum_{k=1}^\infty k
    \left( \frac{e^{\sigma\, i\, \pi\, (\frac 1 2 -g)}\,
    \lambda^{2 g}}{\rho(\alpha,\beta)}\right)^k , \ {\rm for}\
    -\frac 1 2 < g < 0\, ,} \\ \\
           \displaystyle{
    -\frac{2\,g}{\lambda}\,
    \sum_{k=1}^\infty k  \left(\rho(\alpha,\beta)\,
    {e^{-\sigma\, i\, \pi\, (\frac 1 2 -g)} \,
    \lambda^{-2 g}}\right)^k , \ {\rm for}\
    0 < g < \frac 1 2\,  ,}
   \end{array}
   \right.
\end{equation}
which are the term by term derivatives of the corresponding
asymptotic series in Eq.\ (\ref{tau-asymp}).

Collecting these results, we have
\begin{equation}\label{des-asymp-deriv-tau-Tr}\begin{array}{c}
   \partial_\lambda \Big[ \tau(\lambda)\,
  Tr\{G_D(\lambda)-G_N(\lambda)\}\Big]
  \sim \\ \\
   \sim \left\{
  \begin{array}{l}
  \displaystyle{
    2\, g \sum_{k=1}^\infty
    \left( \frac{e^{\sigma\, i\, \pi\, (\frac 1 2 -g)}
    }{\rho(\alpha,\beta)}\right)^k
    \left(2\, g \, k -1\right) \lambda^{2\, g \, k -2}, \ {\rm for}\
    -\frac 1 2 < g < 0\, ,} \\ \\
      \displaystyle{ 2\, g
    \sum_{k=0}^\infty \left(\rho(\alpha,\beta)\,
    {e^{-\sigma\, i\, \pi\, (\frac 1 2 -g)}
    }\right)^k
    \left(2\, g \, k + 1\right) \lambda^{-2\, g \, k -2}
    , \ {\rm for}\
    0 < g < \frac 1 2\, .}
  \end{array}
  \right.
\end{array}
\end{equation}
Notice the $g$-dependent powers of $\lambda$ appearing in these
asymptotic expansions.

\subsection{The $\zeta$ and $\eta$ functions} \label{spectral-functions}

The $\zeta$-function for a general selfadjoint extension of $D$
can be defined, for $\Re(s)>1$, as \cite{Seeley}
\begin{equation}\label{zeta}
  \zeta(s)=- \frac{1}{2\,\pi\,i} \oint_{\mathcal{C}}
  \frac{\lambda^{1-s}}{s-1} \, Tr\left\{{G(\lambda)}^2\right\}
  \, d\lambda\, ,
\end{equation}
where the curve $\mathcal{C}$ encircles counterclockwise the
spectrum of $D$. According to Eq.\ (\ref{TrG2}), we have
\begin{equation}\label{zeta1}
  \zeta(s)= \zeta^D(s)+\frac{1}{2\,\pi\,i} \oint_{\mathcal{C}}
  \frac{\lambda^{1-s}}{s-1} \, \partial_\lambda \Big[ \tau(\lambda)\,
  Tr\{G_D(\lambda)-G_N(\lambda)\}\Big]
  \, d\lambda\, ,
\end{equation}
where $\zeta^D(s)$ is the $\zeta$-function for the $D$-extension.

Since the negative eigenvalues of the selfadjoint extension $D_x^{(\alpha,\beta)}$ are minus the positive eigenvalues
corresponding to extension $D_x^{(\alpha,-\beta)}$ (as
discussed in Section \ref{the-spectrum}), we define a partial
$\zeta$-function by means of a path of integration that encircles the
positive eigenvalues only,
\begin{equation}\label{zeta+}\begin{array}{c}
  \displaystyle{  \zeta_+^{(\alpha,\beta)}(s)=
  \frac{1}{2\,\pi\,i} \int_{-i\,\infty+0}^{i\,\infty+0}
  \frac{\lambda^{1-s}}{s-1} \, Tr\left\{{G(\lambda)}^2\right\}
  \, d\lambda =
    } \\ \\
    \displaystyle{
    = \zeta_+^{D}(s)
    -\frac{1}{2\,\pi\,i} \int_{-i\,\infty+0}^{i\,\infty+0}
  \frac{\lambda^{1-s}}{s-1} \,
  \partial_\lambda \Big[ \tau(\lambda)\,
  Tr\{G_D(\lambda)-G_N(\lambda)\}\Big]
  \, d\lambda
    } \, ,
\end{array}
\end{equation}
where $\zeta_+^{D}(s)$ is the partial $\zeta$-function for the
$D$-extension. Notice that we can write
\begin{equation}\label{zeta+1}\begin{array}{c}
  \displaystyle{\zeta_+^{(\alpha,\beta)}(s)=
    \frac{1}{2\,\pi} \int_{1}^{\infty}
    e^{i\,\frac{\pi}{2}\,(1-s)}\,
  \frac{\mu^{1-s}}{s-1} \,
  Tr\left\{{G(e^{i\,\frac{\pi}{2}}\, \mu)}^2\right\}
  \, d\mu\,  +
  } \\ \\
  \displaystyle{
    +\frac{1}{2\,\pi} \int_{1}^{\infty}
    e^{- i\,\frac{\pi}{2}\,(1-s)}\,
  \frac{\mu^{1-s}}{s-1} \,
  Tr\left\{{G(e^{-i\,\frac{\pi}{2}}\, \mu)}^2\right\}
  \, d\mu + \frac{h(s)}{s-1}\, ,
  }
\end{array}
\end{equation}
where $h(s)$ is an entire function. In order to
determine the poles of $\zeta_+^{(\alpha,\beta)}(s)$, we add and
subtract a partial sum of the asymptotic expansion of $Tr\left\{{G(\lambda)}^2\right\}$
(obtained in Section \ref{Asymptotic-expansion}) to the integrands of the right
hand side of Eq.\ (\ref{zeta+1}).

In this way, we get for the $D$-extension and for real $s>1$
\begin{equation}\label{zetaD+1}\begin{array}{c}
  \displaystyle{ \zeta_+^{D}(s) =
    \frac{1}{2\,\pi\,(s-1)} \int_{1}^{\infty}
    e^{i\,\frac{\pi}{2}\,(1-s)}\,
  {\mu^{1-s}} \,
  \left\{ \sum_{k=2}^{N} e^{-i\,\frac{\pi}{2}\,k}\,
  A_k(g,1) \, \mu^{-k} \right\}
  \, d\mu\,  +
  } \\ \\
  \displaystyle{
    \frac{1}{2\,\pi\,(s-1)} \int_{1}^{\infty}
    e^{-i\,\frac{\pi}{2}\,(1-s)}\,
  {\mu^{1-s}} \,
  \left\{ \sum_{k=2}^{N} e^{i\,\frac{\pi}{2}\,k}\,
  A_k(g,1)^* \, \mu^{-k} \right\}
  \, d\mu + \frac{h_N(s)}{s-1}=
  } \\ \\
  \displaystyle{
  = \frac{1}{\pi\,(s-1)}\,\sum_{k=0}^{N}\frac{1}{s+k}\, \Re \left\{
  e^{-i\,\frac{\pi}{2}\,(s+k+1)}\,A_{k+2}(g,1) \right\}
  + \frac{h_N(s)}{s-1}\, ,
  }
\end{array}
\end{equation}
where $h_N(s)$ is an analytic function in the open half plane
$\Re(s)> 1-N$. In consequence, the meromorphic extension of $\zeta_+^{D}(s)$
presents a simple pole at $s=1$ (see Eq.\ (\ref{zetaD+1})) with a
residue given by (see Eq.\ (\ref{traza-derivGD}))
\begin{equation}\label{residuos=1}
  \left. {\rm Res}\,\zeta_+^{D}(s) \right|_{s=1}
  =  \frac{1}{2\, \pi \, i} \int_{-i\, \infty+0}^{i\, \infty+0}
  \lambda^0 \, \partial_\lambda \left\{
  \frac{J_{g+\frac 1 2}(\lambda)}{J_{g-\frac 1 2}(\lambda)}
  \right\}\, d\lambda = \frac 1 \pi\, .
\end{equation}

It also presents simple poles at $s=-k$, for $k=0,1,2,\dots$, with
residues given by
\begin{equation}\label{otros-residuos}
  \left. {\rm Res}\,\zeta_+^{D}(s) \right|_{s=-k}
  = \frac{\Im\left\{A_{k+2}(g,1)\right\}}{\pi(k+1)}  \, ,
\end{equation}
with the coefficients $A_k(g,1)$ given by Eq.\
(\ref{asymp-trGD-upp}). In particular, notice that these residues
vanish for even $k$.

\bigskip

For a general selfadjoint extension $D_x^{(\alpha,\beta)}$, we
must also consider  the singularities coming from the asymptotic
expansion of $\partial_\lambda [ \tau(\lambda)$
$Tr\{G_D(\lambda)-G_N(\lambda)\}]$ (see Eq.\
(\ref{des-asymp-deriv-tau-Tr})). For definiteness, let us consider
in the following the case $-\frac 1 2 < g <0$ (the case $0<g<\frac
1  2$ leads to similar results).

From Eq.\ (\ref{zeta+}), and taking into account Eq.\
(\ref{zeta+1}), for real $s>1$ we write
\begin{equation}\label{zetadif}\begin{array}{c}
    \zeta_+^{(\alpha,\beta)}(s)-\zeta_+^{D}(s)=
    \displaystyle{\frac{H_N(s)}{s-1}} \, -\\ \\
  \displaystyle{-
     \frac{ g}{\pi\,(s-1)} \int_{1}^{\infty}
    e^{i\,\frac{\pi}{2}\,(-s-1)}\,
  {\mu^{1-s}} \,
  \left\{ \sum_{k=1}^{N}\left(\frac{ e^{i\,\frac{\pi}{2}}}
  {\rho(\alpha,\beta)}\right)^k\,
  (2\,g\,k-1) \, \mu^{2\,g\,k-2} \right\}
  \, d\mu
  } \\ \\
   -\displaystyle{
       \frac{g}{\pi\,(s-1)} \int_{1}^{\infty}
    e^{-i\,\frac{\pi}{2}\,(-s-1)}\,
  {\mu^{1-s}} \,
  \left\{ \sum_{k=1}^{N}\left(\frac{ e^{-i\,\frac{\pi}{2}}}
  {\rho(\alpha,\beta)}\right)^k\,
  (2\,g\,k-1) \, \mu^{2\,g\,k-2} \right\}
  \, d\mu
  } \\ \\
  \displaystyle{
  =- \frac{\,2\,g}{\pi\,(s-1)}\, \sum_{k=1}^{N}\,
  \left(\frac{2\,g\,k-1}{s-2\,g\,k}\right)\, \Re \left\{
  \frac{e^{i\,\frac{\pi}{2} (k-s-1)}}{\rho(\alpha,\beta)^k}\right\}
  + \displaystyle{\frac{H_N(s)}{s-1}}\, ,
  }
\end{array}
\end{equation}
where $H_N(s)$ is holomorphic for $\Re(s)>2\,g\,(N+1)$. We conclude that $\zeta_+^{(\alpha,\beta)}(s)-\zeta_+^{D}(s)$ has a
meromorphic extension which presents a simple pole at $s=1$, with
a vanishing residue,
\begin{equation}\label{res-dif-s=1}\begin{array}{c}
    \left. {\rm Res}\,\left(\zeta_+^{(\alpha,\beta)}(s)-
  \zeta_+^{D}(s)\right) \right|_{s=1}
  = \\ \\ \displaystyle{
  = -\frac{1}{2\,\pi\,i} \int_{-i\,\infty+0}^{i\,\infty+0}
  {\lambda^{0}}\,
  \partial_\lambda \Big[ \tau(\lambda)\,
  Tr\{G_D(\lambda)-G_N(\lambda)\}\Big]
  \, d\lambda =0}\, ,
\end{array}
\end{equation}
as follows from Eqs.\ (\ref{trdifasymp}) and (\ref{tau-asymp}).

Notice also the presence of simple poles located at negative non
integer $g$-dependent positions $s=2gk$ for
$k=1,2,\dots$, with residues which also depend on the selfadjoint
extension, given by
\begin{equation}\label{res-g-dep}
      \left. {\rm Res}\,\left\{\zeta_+^{(\alpha,\beta)}(s)-
  \zeta_+^{D}(s)\right\} \right|_{s=2\,g\,k} =
   \frac{-\, 2\,g}{\pi\,\rho(\alpha,\beta)^k}
    \ {\sin\left[\left(\frac{1}{2} -g\right)k\,\pi\right]}
  \, .
\end{equation}

Following the discussion after Eq.\
(\ref{eigenvalues-neg}), we get for the complete $\zeta$-function
\begin{equation}\label{zeta-completa}
  \zeta^{(\alpha,\beta)}(s) = \zeta_+^{(\alpha,\beta)}(s)+
  e^{-i\,\pi\,s} \, \zeta_+^{(\alpha,-\beta)}(s)\, .
\end{equation}
In particular, for the $D$-extension -whose spectrum is symmetric with respect to
the origin (see Eq.\ (\ref{alpha0}))- we get
\begin{equation}\label{zeta-alpha0}
  \zeta^D(s) = \left(1 + e^{-i\,\pi\,s}\right)
  \zeta_+^{D}(s)\,.
\end{equation}
Notice that $\zeta^D(s)$ is an entire function. Indeed, from Eq.\
(\ref{otros-residuos}), the residue at $s=-k$ vanishes for $k$
even, whereas for $k=2\,l+1$, with $l=0,1,2,\dots$, we get
\begin{equation}\label{residuos-nulos}
   \left. {\rm Res}\,\left\{\zeta^{D}(s)\right\}
   \right|_{s=-2l-1} =
   \left(1 + e^{i\,\pi\left(2l+1\right)}\right)
   \left. {\rm Res}\,\left\{\zeta_+^{D}(s)\right\}
   \right|_{s=1-2\,l} =0\, .
\end{equation}

On the other hand,  for a general selfadjoint extension, the
singularities of $\zeta^{(\alpha,\beta)}(s)$ are simple poles
located at $s=2\,g\,k<0$, for $k=1,2,\dots$, with residues
\begin{equation}\label{residues-zeta-completa}\begin{array}{c}
 \displaystyle{
  \left. {\rm Res}\,\left\{\zeta^{(\alpha,\beta)}(s)-
  \zeta^{D}(s)\right\} \right|_{s=2\,g\,k} =
  }\\ \\
  = \displaystyle{\left. {\rm Res}\,\left\{
  \left[\zeta_+^{(\alpha,\beta)}(s)-
  \zeta_+^{D}(s)\right] + e^{-i\,\pi\,s} \,
  \left[\zeta_+^{(\alpha,-\beta)}(s)-
  \zeta_+^{D}(s)\right]
  \right\} \right|_{s=2\,g\,k} } =\\ \\
  \displaystyle{=
  (-1)^k \,
  \frac{2\,g}{\pi} \
  \frac{\sin(2\,g\,k\,\pi)}{\rho(\alpha,\beta)^k}}\,
  e^{i\,\pi\left(\frac 1 2 -g\right) k}\, ,
\end{array}
\end{equation}
where we have used $\rho(\alpha,-\beta)=-\rho(\alpha,\beta)$ (see
Eq.\ (\ref{rho})).

\bigskip

Similarly, for the spectral asymmetry \cite{Atiyah:1975jf} we have
\begin{equation}\label{eta-completa}
  \eta^{(\alpha,\beta)}(s) = \zeta_+^{(\alpha,\beta)}(s)-
  \zeta_+^{(\alpha,-\beta)}(s)\, .
\end{equation}
In particular, $\eta^{(0,1)}(s)= \eta^{(1,0)}(s)\equiv 0 $,
since the corresponding spectra are symmetric (see eqs.\ (\ref{alpha0}) and
(\ref{eigen-beta0})). For a general selfadjoint extension and $-\frac 1 2 < g <0$ the function
$\eta^{(\alpha,\beta)}(s)$ presents simple poles at $s=2\,g\,k$,
for odd $k=1,3,5,\dots$, with residues given by
\begin{equation}\label{residues-eta}
\displaystyle{
  \left. {\rm Res}\,\left\{\eta^{(\alpha,\beta)}(s)\right\}
  \right|_{s=2\,g\,k} = -
  \frac{4\,g}{\pi} \,
  \frac{\sin\left[\left(\frac{1}{2}-g\right)
  \,k\,\pi\right]}{\rho(\alpha,\beta)^k}}\, .
\end{equation}

For the case $0<g<\frac 1 2$, an entirely similar calculation
shows that $\zeta_+^{(\alpha,\beta)}(s)-\zeta_+^{D}(s)$ has a
meromorphic extension which presents simple poles at negative non
integer $g$-dependent positions, $s=-2\,g\,k$, for $k=1,2,\dots$,
with residues depending on the selfadjoint extension, given by
\begin{equation}\label{res-gpos-dep}\begin{array}{c}
      \left. {\rm Res}\,\left\{\zeta_+^{(\alpha,\beta)}(s)-
  \zeta_+^{D}(s)\right\} \right|_{s=-2\,g\,k} =
    \\ \\
    = \displaystyle{ -\,\frac{2\,g}{\pi}\,\rho(\alpha,\beta)^k
    \ {\sin\left[\left(\frac{1}{2} -g\right)k\,\pi\right]}
  }\, .
\end{array}
\end{equation}
From this result, it is immediate to get the residues for the
$\zeta$- and $\eta$-functions. One gets the same expressions as in
the right hand sides of eqs.\ (\ref{residues-zeta-completa}) and
(\ref{residues-eta}), with $\rho(\alpha,\beta)$ and
$e^{i\,\pi\left(\frac 1 2 -g\right) k}$ replaced by their
inverses.

\subsection{Scale Invariance}

Let us remark that when neither $\alpha$ nor $\beta$ is 0, the
residue of $\zeta_+^{(\alpha,\beta)}$ at $s=-2\,|g|\,k$ is a
constant times $(\beta/\alpha)^{k\,{\rm sign}(g)}$. This is
consistent with the behavior of $D$ under the scaling isometry
$T_L\,u(x):=L^{-1/2}\,u(x/L)$ mapping $\mathbf{L_2}(0,1)$ onto $\mathbf{L_2}(0,L)$. The extension $D^{(\alpha,\beta)}$ is
unitarily equivalent to the operator
$L\,\dot{D}^{(\dot{\alpha},\dot{\beta})}$ defined on
$\mathbf{L_2}(0,L)$, with $\dot{\alpha} := L^{g}\, \alpha$ and
$\dot{\beta}: = L^{-g} \, \beta$:
\begin{equation}\label{isometry}
  T_L\,D^{(\alpha,\beta)} := L\,
  \dot{D}^{(\dot\alpha,\dot\beta)}\, T_L\, .
\end{equation}
Notice that only for the extensions with $\alpha=0$ or $\beta=0$
the boundary condition at the singular point $x=0$ -given by Eq.\
(\ref{BC2})- is scale invariant.

The partial $\zeta$-function of the scaled operator can be written as
\begin{equation}\label{zetas-isometry}
  \dot{\zeta}_+^{(\dot\alpha,\dot\beta)}(s)=
   L^{s}\,\zeta_+^{(\alpha,\beta)}(s)\, ,
\end{equation}
whereas its residues satisfy
\begin{equation}\label{zetas-isometry-residues}
   \left. {\rm Res}\,\left\{\dot{\zeta}_+^{(\dot\alpha,\dot\beta)}(s)
   \right\} \right|_{s=-2\,|g|\,k} = L^{-2\,|g|\,k}
   \left. {\rm Res}\,\left\{{\zeta}_+^{(\alpha,\beta)}(s)
   \right\} \right|_{s=-2\,|g|\,k}\, .
\end{equation}
The factor $L^{-2\,|g|\,k}$ exactly cancels the effect the change
in the boundary condition at the singularity has on
$\rho(\alpha,\beta)$,
\begin{equation}\label{change-in-rho}
  \rho(\alpha,\beta)^{k\,{\rm sign}(g)}
  =L^{2\,|g|\,k}\, \rho(\dot\alpha,\dot\beta)^{k\,{\rm sign}(g)}\, .
\end{equation}
Thus the length of the intervals $(0,1)$ and $(0,L)$
has no effect on the structure of these residues, which presumably
are determined locally in a neighborhood of $x=0$.

\bigskip

Finally, let us point out that these anomalous poles are not
present in the $g=0$ case. Indeed, in this case $\tau(\lambda)$ in
Eq.\ (\ref{taudelambda}) has a constant asymptotic expansion,
whereas $Tr\{G_D(\lambda)-G_N(\lambda)\} \sim 0$ (see Eq.\
(\ref{trdifasymp})). Moreover, the residues of the poles coming
from $\zeta_+^{D}(s)$ vanish (see Eqs.\
(\ref{otros-residuos}) and (\ref{asymp-trGD-upp})), except for the
one at $s=1$, whose residue is $1/\pi$ (see Eq.\ (\ref{residuos=1})).

Consequently, the presence of poles in the spectral functions
located at non integer positions is a consequence of the singular
behavior of the 0-th order term in $D$ near the origin, together
with a boundary condition which breaks the scale invariance.

\section{The second order case}\label{second-order}

In this section we will show that one obtains similar results for the selfadjoint extensions of the second order differential
operator
\begin{equation}\label{Delta}
  H = - \partial_x^2 + \frac{g(g-1)}{x^2}\, ,
\end{equation}
defined on a set of functions $\phi(x)\in \mathcal{C}_0^\infty(0,1)$. Since we are interested in the effects of the singularity at $x=0$, for definiteness we impose again $\phi(1)=0$. Proceeding as in the previous section \cite{Falomir:2004zf}, for $|g| < \frac 1 2$ (to be congruent with the first-orden case analyzed in Section \ref{the-operator}), one can show that $H$ admits a continuous family of selfadjoint extensions in $\mathbf{L}_2\left(0,1\right)$, $H^{(\alpha,\beta)}$, characterized by two real parameters $\alpha,\beta$ satisfying $\alpha^2+\beta^2=1$ and defined on functions which behave near the origin as
\begin{equation}\label{BC-second}
  \phi(x)= C_1 \, x^g + C_2\, x^{1-g} + O(x^{3/2}) \, ,
\end{equation}
where the coefficients $C_{1},C_{2}$ satisfy
\begin{equation}\label{BC21}
    \alpha\, C_1+ \beta\, C_2 = 0\, .
\end{equation}
The spectrum of $H^{(\alpha,\beta)}$ is determined by the relation (analogous to
Eq.\ (\ref{eigenvalues-pos}))
\begin{equation}\label{spectra-second}
  \mathcal{F}(\mu):= \frac 1 \mu\, F(\mu) = \varrho (\alpha,\beta)\, ,
\end{equation}
where now we define
\begin{equation}\label{varrho}
  \varrho(\alpha,\beta) := \left(\frac \beta \alpha \right)
  2^{2\,g-1} \, \frac{\Gamma(\frac 1 2 +g)}
  {\Gamma(\frac 3 2 -g)} \, .
\end{equation}

Also in this case,  $\alpha = 0$ and $\beta =0$ correspond to two
scale invariant boundary conditions at the singularity. For these
two limiting extensions, it is easily seen from Eqs.\
(\ref{ec-dif-g-diag}), (\ref{near0-D}) and (\ref{near0-N}) that
the resolvent of $H^{(\alpha,\beta)}$ satisfies
\begin{equation}\label{resolv-second}
  \mathcal{G}^{D,N}(x,y;\mu^2) =
  \frac 1 \mu \, G_{11}^{D,N}(x,y;\mu)\, .
\end{equation}

The resolvent for a general selfadjoint extension
$H^{(\alpha,\beta)}$ is constructed as a convex linear
combination of $\mathcal{G}^{D}(\mu^2)$ and
$\mathcal{G}^{N}(\mu^2)$ as in (\ref{linear-comb}), with a
coefficient
\begin{equation}\label{tau-second}
  \tau(\mu) =  \displaystyle{
  \frac{1}{1-
  \displaystyle{
  \frac{\varrho(\alpha,\beta)}{\mathcal{F}(\mu)}} } }\, .
\end{equation}

Following the methods employed for the first order case \cite{Falomir:2004zf}, one can show that the $\zeta$-function associated to $H^{(\alpha,\beta)}$ has a meromorphic extension which presents simple poles located
at negative $g$-dependent positions,
\begin{equation}\label{polos-zeta-orden2}
    s_k=-\left( \frac 1 2 -g \right) k\, ,
     \quad {\rm for}\  k=1,2,\dots\, ,
\end{equation}
with residues which depend on the SAE given by
\begin{equation}\label{res-g-dep-orden2}\begin{array}{c}
      \left. {\rm Res}\,\left\{\zeta^{(\alpha,\beta)}(s)-
  \zeta^{D}(s)\right\} \right|_{s=s_k} =
    \\ \\
    = \displaystyle{-\left( \frac{2\,g-1}{2\, \pi}\right)
    \varrho(\alpha,\beta)^k
    \ {\sin\left[\frac{\pi}{2}\left( 2g-1\right)k \right]}
  }\, .
\end{array}
\end{equation}
Notice that these poles are irrational for irrational values of $g$. Moreover, these residues vanish for the ``N-extension" ($ \varrho(\alpha,0)=0$), and have a singular limit for $\alpha\rightarrow 0$. As in the first order case, this behavior can also be explained through scaling arguments \cite{Falomir:2004zf}.

Finally, let us remark that the relation between the
$\zeta$-function and the trace of the heat-kernel of
$D_x^{(\alpha,\beta)}$ (See Appendix \ref{spectralfunctionsrelations}) straightforwardly lead to the
following small-$t$ asymptotic expansion,
\begin{equation}\label{heat-asymp-orden2}\begin{array}{c}
  \displaystyle{  Tr\left\{e^{-t\,D_x^{(\alpha,\beta)}}-
  e^{-t\,D_x^{D}}\right\}
  \sim  \left(g-\frac 1 2\right) -} \\ \\
  \displaystyle{- \sum_{k=1}^\infty}
  \left\{\Gamma\left(\left[\frac 1 2-g \right]k\right)
  \frac{2\,g-1}{2\, \pi}\,\varrho(\alpha,\beta)^k
    \ {\sin\left[
    \frac{\pi}{2}\left( 2g-1\right)k \right]}\right\}
    t^{\left(\frac{1}{2}-g\right)k}\, .
\end{array}
\end{equation}
The first term in the right hand side coincides with the
result reported in \cite{Mooers}. Notice also the $g$-dependent
powers of $t$ appearing in the asymptotic series in the right hand
side of Eq.\ (\ref{heat-asymp-orden2}) for the general SAE and the dependence of the SDW coefficients on the $(\alpha,\beta)$-parameters. In particular, the first term in this series
reduces to
\begin{equation}\label{first-term}
    - \frac{\beta}{\alpha}\, \frac{2^{2g-1}}{\Gamma(\frac 1 2 -g)}\
     t^{\frac 1 2 -g}\, .
\end{equation}
This power of $t$ also coincides with the result quoted in
\cite{Mooers}, but we find a different coefficient.

\section{The $g=\frac 1 2$  case}\label{limitcase}

The second order differential operator given by Eq.\ (\ref{Delta}) with $g=\frac 1 2$,
\begin{equation}\label{g1/2}
    H_0=-\frac{d^2}{dx^2}-\frac{1}{4\, x^2}\, ,
\end{equation}
defined on the domain $\phi(x)\in\mathcal{C}_0^\infty(0,1)$, also admits a continuous family of nontrivial SAE in $\mathbf{L}_2(0,1)$, which reflects in even more unusual properties of its spectral functions. This problem has been considered by K.\ Kirsten et al.\  in \cite{Kirsten:2005bh,Kirsten:2005yw,Kirsten:2007ur,Kirsten:2008wu}, articles where an error in Appendix A of \cite{Falomir:2004zf} has been corrected.

Also here, for definiteness, we impose $\phi(1)=0$. As before, the SAE $H_0^{(\alpha,\beta)}$ are characterized by two real parameters satisfying $\alpha^2+\beta^2=1$ and defined on a domain of functions which behave near the origin as
\begin{equation}\label{BC-second-lim}
  \phi(x)= C_1 \, \sqrt{x} + C_2\, \sqrt{x}\log{x} + O(x^{3/2}) \, ,
\end{equation}
where the coefficients $C_{1},C_{2}$ satisfy Eq.\ (\ref{BC21}).

The eigenfunction of $H_0^D:=H_0^{(0,1)}$ corresponding to an eigenvalue
$\lambda=\mu^2$ is given by,
\begin{equation}
    \phi(x)=\sqrt{x}\, J_0(\mu x)\,.
\end{equation}
The condition $\phi(1)=0$ tells that $\mu$ is a (positive) zero of the Bessel function $J_0(z)$. On the other hand, for an arbitrary SAE $H_0^{(\alpha,\beta)}$ with $\alpha\neq 0$, the eigenfunction corresponding to an eigenvalue $\lambda=\mu^2$ is given by
\begin{equation}
  \phi(x)=\left\{C_1-C_2(\log{\mu/2}+\gamma)
    \right\}\,\sqrt{x}\,J_0(\mu x)\,
    +
  \frac{\pi}{2}\, C_2\,\sqrt{x}\,N_0(\mu x) \, ,
\end{equation}
where $C_1,C_2$ are constrained by Eq.\ (\ref{BC21}). The condition $\phi(1)=0$ leads to the equation
\begin{equation}
    2(\theta-\log{\mu})J_0(\mu)+
    \pi N_0(\mu)=0\, ,
\end{equation}
where $\theta:=-\beta/\alpha+\log{2}-\gamma$, which determines the
spectrum of $H_0^{(\alpha,\beta)}$. Notice that there are no
negative eigenvalues.

\bigskip

The trace of the resolvent $\mathcal{G}_0^D(\mu^2):=(H_0^{D}-\mu^2)^{-1}$ can be readily computed to get
\begin{equation}\label{traces-00}
  {\rm Tr}\,\left\{\mathcal{G}_0^{D}(\mu^2)\right\}=\frac{1}{2\mu}\frac{J_1(\mu)}{J_0(\mu)} \, .
\end{equation}
This trace admits the following asymptotic expansion in integer powers of $\mu$
\begin{eqnarray}\label{tr-gd-0}
      Tr\left\{\mathcal{G}^{D}(\mu^2)\right\}\sim
    \frac{e^{i\sigma \frac \pi 2}}{2\mu} \left(
    \frac{P(1,\mu)- i \sigma \, Q(1,\mu)}
    {P(0,\mu)- i \sigma \, Q(0,\mu)}\right)
    \sim
    \\\nonumber
    \sim{\frac{i\,\sigma }{2\mu } } +
  \frac{1}{4\,{\mu }^2} +
  {\frac{i \,\sigma}{16{\mu }^3} } -
  \frac{1}{16\,{\mu }^4} +
  {{O}({\mu^{-5} })} \, ,
\end{eqnarray}
where $\sigma = +1$ ($-1$) for $\Im(\mu)>0$ ($\Im(\mu)<0$). From this asymptotic expansion one concludes that the poles of the corresponding $\zeta$-function ${\rm Tr}\,\left\{H^D_0\right\}^{-s}$ are located at $s=1/2-k$, for $k=0,1,2,\dots$

On the other hand, the trace of the resolvent $\mathcal{G}_0^{(\alpha,\beta)}(\mu^2):=(H_0^{(\alpha,\beta)}-\mu^2)^{-1}$, corresponding to a general selfadjoint extension, gives
\begin{eqnarray}\label{traces-0}
  {\rm Tr}\,\left\{\mathcal{G}_0^{(\alpha,\beta)}(\mu^2)\right\}
    = \frac{1}{2\mu}\, \frac{2(\theta-\log{\mu})J_1(\mu)+\pi N_1(\mu)}
    {2(\theta-\log{\mu})J_0(\mu)+\pi N_0(\mu)}+\\\nonumber
    \mbox{}+\frac{J_0(\mu)}{\mu^2[\pi N_0(\mu) + 2 (\theta - \log\mu)]
    J_0(\mu)}
    \, .
\end{eqnarray}
The asymptotic expansion of the first term in the {\small R.H.S.}\ of Eq.\ (\ref{traces-0}) is also given by expression (\ref{tr-gd-0}). Therefore, it leads to the standard poles of the corresponding $\zeta$-function, ${\rm Tr}\,\left\{H^{(\alpha,\beta)}_0\right\}^{-s}$, located at $s=1/2-k$, for $k=0,1,2,\dots$ However, the asymptotic expansion of the second term in the {\small R.H.S.}\ of Eq.\ (\ref{traces-0}) is given by
\begin{equation}\label{missing-asymp}
    \frac{J_0(\mu)}{\mu^2[\pi N_0(\mu) + 2 (\theta - \log\mu)J_0(\mu)]}
    \sim \frac{1}{\mu^2[i \pi \sigma + 2 (\theta - \log \mu)]} \,,
\end{equation}
with $\sigma =1$ ($\sigma =-1$) for $\mu$ in the upper
(lower) half-plane. This expression gives an additional contribution to the $\zeta$-function  given by
\begin{equation}\label{contrib-adic-zeta1}
\begin{array}{c} \displaystyle
    {\rm Tr}\,\left\{H^{(\alpha,\beta)}_0\right\}^{-s}-
    {\rm Tr}\,\left\{H^D_0\right\}^{-s}=
  \\ \\ \displaystyle
  = {\frac{ e^{- 2\,\theta s }\,
         }{2\,\pi \,i}} \left[ e^{ i  \,\pi \,s}\,
         {\Gamma}\left(0,
           \left( \frac{i \, \pi}{2}  -
              2  \,\theta  \right) s
             \right) -e^{ -i  \,\pi \,s}\, {\Gamma}\left(0,\left(
         \frac{-i \,\pi}{2} - 2\,\theta \right) s\right)
        \right]=
        \\ \\ \displaystyle
        =-\frac{1}{\pi}\, e^{ 2  s \left( \frac{\beta}{\alpha}-\log{2}+\gamma \right)}
        \, {\sin(\pi s)}\, \log s + H(s) \,,
\end{array}
\end{equation}
where $H(s)$ is an entire function of $s$. Therefore, the $\zeta$-function of the general SAE of $H_0$ develops a cut on the negative real axis, which is present even for $\beta=0$  and has a singular behavior for $\alpha\rightarrow 0$.

As a consequence, the derivative of the $\zeta$-function in a neighborhood of the origin behaves as \cite{Kirsten:2005bh,Kirsten:2005yw}
\begin{equation}\label{around0-deriv}
    \frac{d}{ds}{{\rm Tr}\,\left\{H^{(\alpha,\beta)}_0\right\}^{-s}}=
    -\log s + O(1)\, .
\end{equation}
This logarithmic branch cut from the origin for the general SAE makes the functional determinant in Eq.\ (\ref{alazeta}) ill-defined (See the discussions in \cite{Kirsten:2005bh,Kirsten:2005yw}). Notice that this logarithmic cut is absent only for the unique SAE of $H_0$ which is
locally scale invariant, namely $H^D_0=H_0^{(0,1)}$.

On the other hand, the term (\ref{missing-asymp}) gives the following contribution to the heat-kernel trace of a
general SAE of $H_0$
\begin{equation}\label{heat-kernel0}
    \begin{array}{c} \displaystyle
      {\rm Tr}\left\{e^{-tH^{(\alpha,\beta)}_0 }\right\} -
      {\rm Tr}\left\{e^{-tH^{(\alpha,\beta)}_0 }\right\} =
      \\ \\ \displaystyle
      = -\oint \frac{d\mu}{\pi i\mu}\, e^{-t \mu^2} \frac{J_0(\mu)}{\pi N_0(\mu) + 2 (\theta - \log\mu)J_0(\mu)}
    =
    \\ \\ \displaystyle
    =\frac{1}{\pi} \, \Im \int_1^\infty \frac{e^{-t \, x}}
    {x\left(\log x -2 \theta +i \pi\right)}\, d x
    + R(t)
    \, ,
    \end{array}
\end{equation}
where $R(t)$ is a smooth function at $t=0$. It was shown in \cite{Kirsten:2005bh} that the integral in the right hand side of Eq.\ (\ref{heat-kernel0}) has an asymptotic expansion for small
$t$ in terms of negative integer powers of $\log t$.

\section{Non-compact case}\label{adjoint-H}

In this Section we will consider a locally homogeneous (near a singularity) symmetric second order differential operator on the noncompact one-dimensional manifold $M=\mathbb{R}^+$. We will see that also in this case the associated $\zeta$-function presents a non-standard singularity structure, related with the breaking of this scale homogeneity by the definition domains of the SAE of this operator. We will employ a different approach to this problem, based on the von Neumann's theory of self adjoint extensions of symmetric operators. This allows to express the spectrum of the SAE in terms of a trascendental equation from which we are able to derive an asymptotic expansion of the eigenvalues. This eventually leads to the pole structure of the $\zeta$-function we are interested in.

Then, following \cite{FPW}, let us consider the operator
\begin{equation}\label{ham}
    H=-\frac{d^2}{dx^2}+V(x),
\end{equation}
with
\begin{equation}
    V(x)=\frac{\nu^2-\frac{1}{4}}{x^2}+x^2,
\end{equation}
densely defined on the domain ${ \mathcal D}(H)={\mathcal
C}_0^\infty(\mathbb{R}^+)$, the linear space of smooth functions
$\phi(x)$ with $x\in\mathbb{R}^+$ and compact support out of the origin. We have added to the singular at the origin term of the potential an $x^2$ term in order to get discrete spectra.

According to von Neumann's theory of deficiency indices \cite{Reed-Simon}, in order to get the SAE of $H$ in $\mathbf{L}_2\left(\mathbb{R}^+\right)$ we need to compute the adjoint $H^\dagger$ and determine the deficiency subspaces ${\mathcal K}_{\pm}:={\rm Kernel}\,(H^\dagger\mp i)$. The domain of $H^\dagger$ is the subspace
of square integrable functions having an absolutely continuous
first derivative and such that
\begin{equation}\label{Hmas}
  H^\dagger\psi(x)= -\psi''(x)+V(x) \psi(x)\in
\mathbf{L_2}(\mathbb{R}^+)\,.
\end{equation}
Notice that no boundary condition is imposed at $x=0$. To compute the deficiency indices $n_\pm := {\rm dim}\,{\mathcal K}_{\pm}$ of $H$ we must solve the
eigenvalue problem
\begin{equation}\label{eigequ}
    H^\dagger\phi_{\lambda}=-\phi_\lambda''(x)+
    V(x) \phi_\lambda(x)=\lambda \phi_{\lambda},
\end{equation}
for $\phi_{\lambda}\in {\mathcal D}(H^\dagger)$ and
$\lambda\in\mathbb{C}$ with imaginary part $\Im(\lambda)\neq
0$. Let us define the parameter
\begin{equation}
    \alpha:=\frac{1}{2}+\nu\,.
\end{equation}
For the case $0 \leq \nu <1$, i.e.\ $1/2\leq\alpha<3/2$, for any $\lambda\in\mathbb{C}$  Eq.\ (\ref{eigequ}) has a unique nontrivial square-integrable solutions given by
\begin{equation}\label{eigfun}
    \phi_{\lambda}(x)=x^{\alpha}\, e^{-\frac{x^2}{2}}\,
    U\left(\frac{2\alpha+1-\lambda}{4};
    \alpha+\frac{1}{2};x^2\right)\,,
\end{equation}
where $U$ is the confluent hypergeometric function as defined in \cite{A-S}. Thus, the deficiency subspaces
${\mathcal K}_\pm$ are one-dimensional, $n_\pm=1$, and $H$ admits a one-parameter family of SAE\footnote{This is in accordance to Weyl's criterion
\cite{Reed-Simon} according to which, for continuous $V(x)$, $H$
is essentially selfadjoint if and only if it is in the limit
point case, both at infinity and at the origin.

In addition, if $V(x)\geq M>0$, for $x$ large enough, then $H$ is
in the limit point case at infinity. In consequence, in the
present case $H$ is essentially selfadjoint if and only if it is
in the limit point case at zero.

In particular, for positive $V(x)$ (\emph{i.e.} $\nu^2\geq \frac{1}{4}$), if $V(x)\geq \frac{3}{4}\,
x^{-2}$ for $x$ sufficiently close to zero then $H$ is in the
limit point case at the origin. On the contrary, if $V(x)\leq
(\frac{3}{4}-\eps) x^{-2}$, for some $\eps>0$, then $H$ is in the limit
circle case at zero.} which are
in a one-to-one correspondence with the isometries from ${\mathcal
K}_+$ onto $\mathcal{K}_-$. The deficiency subspaces ${\mathcal K}_+$ and ${\mathcal K}_-$
are spanned by $\phi_+:=\phi_{\lambda=i}$ and $\phi_-:=\phi_{\lambda=-i}=\phi_{+}^*$, respectively. Each
isometry ${\mathcal U}_{\gamma}:{\mathcal K}_+\rightarrow
{\mathcal K}_-$ can be characterized by a parameter $\gamma\in
[0,\pi)$ defined by
\begin{equation}\label{isom}
    {\mathcal U}_{\gamma}
    \phi_+ = e^{-2i\gamma}\phi_-\,.
\end{equation}
Each isometry is identified with a selfadjoint extension $H_\gamma$, a closed restriction of $H^{\dagger}$ to a linear subspace
\begin{equation}
    {\mathcal D}(H_{\gamma})\subset
    {\mathcal D}(H^{\dagger})={\mathcal D}(\overline{H})\oplus
    {\mathcal K}_+\oplus {\mathcal K}_- \,,
\end{equation}
where $\overline{H}$ is the closure of $H$. The domain $\mathcal{D}(H_\gamma)$ is defined as the set of functions that can be written as
\begin{equation}\label{phi-suma}
    \phi=\phi_0+A \left( \phi_+ + e^{-2i\gamma}\phi_- \right),
\end{equation}
with $\phi_0\in {\mathcal D}(\overline{H})$ and $A$ a complex constant, and the action of the operator $H_{\gamma}$ by
\begin{equation}\label{definiciondelaSAE}
    H_{\gamma} \phi=H^{\dagger}\phi= H^{\dagger}\phi_0 + \imath  A \left( \phi_+ - e^{-2i\gamma}\phi_- \right)\,.
\end{equation}

Let us consider, for simplicity, the repulsive case ($\nu\geq \frac{1}{2}$), that is $1\leq \alpha <3/2$. In Appendix
\ref{closure} we show that  $\phi_0(x)=o(x^{\alpha})$ and
$\phi_0'(x)=o(x^{\alpha-1})$. Therefore, from Eq.\  (\ref{phi-suma}) and Eq.\ (\ref{eigfun}) with $\lambda = \pm \imath$ we get the following condition for $\phi(x)\in \mathcal{D}(H_\gamma)$ near the origin\footnote{\label{dosLSInv}Notice that
\begin{equation}\label{fimas}
    \phi_+(x)=x^{1-\alpha } \left(\frac{\Gamma \left(\alpha
   -\frac{1}{2}\right) }{\Gamma \left(\frac{1}{4} (2   \alpha
   +(1-i))\right)}+O\left(x^2\right)\right)+x^{\alpha }
   \left(\frac{\Gamma \left(\frac{1}{2}-\alpha
   \right)}{\Gamma
   \left(\left(\frac{3}{4}-\frac{i}{4}\right)-\frac{\alpha }{2}
   \right)}+O\left(x^2\right)\right)\,.
\end{equation}
Then, it is easy to see that there are two locally scale invariant SAE for which the functions in their domains behave near the origin as $x^\alpha$ and $x^{1-\alpha}$ respectively. For any other SAE, the simultaneous presence of both powers of $x$ in the boundary condition near the singularity breaks this local scale homogeneity.},
\begin{equation}\label{bc}
   \partial_x\log{\phi(x)}=\frac{1-\alpha}{x}-
   2\,\frac{\Gamma (\frac{3}{2}-\alpha)}
   {\Gamma (\alpha-\frac{1}{2})}\,
   \frac{\cos{(\gamma-\gamma_1)}}
   {\cos{(\gamma-\gamma_2)}}\cdot x^{2\alpha-2}+
   o(x^{2\alpha-2})\,,
\end{equation}
where we have defined
$\gamma_1=\arg\left\{\Gamma[(-2\alpha+3-i)/4]\right\}$ and
$\gamma_2=\arg\left\{\Gamma[(2\alpha+1-i)/4]\right\}$.

The \emph{boundary condition} specified in Eq.\ (\ref{bc}) characterizes
the domain of the selfadjoint extension $H_\gamma$. In
order to determine its spectrum, we select from the set of eigenfunctions of $H^\dagger$ given in Eq.\ (\ref{eigfun}) those which satisfy Eq.\ (\ref{bc}). This leads to the following transcendental equation for the eigenvalues $\lambda$ \cite{FPW},
\begin{equation}\label{spe}
    \frac{\Gamma \left(\kappa-\frac{\lambda}{4}\right)}
    {\Gamma\left(1-\kappa-\frac{\lambda}{4}\right)}=
    \beta(\gamma,\kappa)\,,
\end{equation}
where we have defined the constants
\begin{equation}\label{param}
  \begin{array}{c}
  \displaystyle{
    \kappa:=\frac{2\alpha+1}{4}\in [3/4,1) }\,,\\ \\
    \displaystyle{
    \beta(\gamma,\kappa):=\cos{(\gamma-\gamma_1)}/\cos{(\gamma-\gamma_2)}}\,.
  \end{array}
\end{equation}
Eq.\ (\ref{spe}) determines the discrete spectrum of the selfadjoint extension characterized by the parameter $\gamma$. From now on we will refer to the SAE as $H_{(\beta)}$, identifying it by the value of $\beta\in \mathbb{R}\cup \{-\infty\}$ defined above.

As expected, the spectrum of $H_{(\beta)}$ is bounded from below; however it presents
a negative eigenvalue for those selfadjoint extensions characterized by
$\beta>\Gamma(\kappa)/\Gamma(1-\kappa)$ (even though the
potential $V(x)\geq 0$). Moreover, there is no
common lower bound; instead, any negative real number is in the spectrum
of some selfadjoint extension.

For any value of $\nu\in [\frac{1}{2},1)$, there are two particular selfadjoint extensions for which the
spectrum can be easily worked out (see Eq.\ (\ref{spe})):
\begin{itemize}
\item {For $\beta=0$  the spectrum is given by
\begin{equation}\label{beta=0}
    \lambda_n=4(n+1-\kappa)\,,\quad n=0,1,2,\ldots
\end{equation}}
  \item {For $\beta= - \infty$  the spectrum
is given by
\begin{equation}\label{beta=-infty}
    \lambda_n=4(n+\kappa)\,,\quad n=0,1,2,\ldots
\end{equation}}
\end{itemize}
For any other value of $\beta$, the eigenvalues also grow linearly with $n$,
\begin{equation}\label{linear-n}
  4(n-1+\kappa)< \lambda_{n}<4(n+\kappa)\,.
\end{equation}

\subsection{Pole structure of the $\zeta$-function }
\label{integ-rep}

We will now study the pole structure of $\zeta_\beta(s)$ corresponding to an arbitrary SAE $H_{(\beta)}$.
Notice that, since the eigenvalues grow linearly with $n$ (see
Eq.\ (\ref{linear-n})), $\zeta_\beta(s)$ is analytic in the open
half-plane $\Re(s)>1$.

Let us begin by considering the $\zeta$-functions for the SAE characterized by $\beta=0$ and
$\beta=-\infty$, which can be explicitly evaluated from the expression of its spectra in Eqs.\ (\ref{beta=0}) and
(\ref{beta=-infty}). We obtain
\begin{eqnarray}\label{zetas}
    \zeta_{\beta=0}(s)=4^{-s} \sum_{n=0}^\infty
    (n+1-\kappa)^{-s}=4^{-s}\zeta(s,1-\kappa)\,,\\
    \zeta_{\beta=-\infty}(s)=4^{-s} \sum_{n=0}^\infty
    (n+\kappa)^{-s}=4^{-s}\zeta(s,\kappa)\,.
\end{eqnarray}
where $\zeta(s,q)$ is the Hurwitz $\zeta$-function, whose
analytic extension presents a unique simple pole at $s=1$ with residue Res\,$\zeta(s,q)|_{s=1}=1$. Therefore, for $\beta=0$ and $\beta=-\infty$ the $\zeta$-function presents a unique simple pole at $s=1$, with residue $1/4$.

For finite $\beta$, let us  define the holomorphic function,
\begin{equation}
    f(\lambda)=\frac{1}
    {\Gamma\left(1-\kappa-\frac{\lambda}{4}\right)}-\frac
    {\beta}{\Gamma \left(\kappa-\frac{\lambda}{4}\right)}\,;
\end{equation}
recall that we are considering the repulsive case $\frac{3}{4}\leq \kappa<1$. The eigenvalues of the
selfadjoint operator $H_{(\beta)}$ correspond to the  zeroes of
$f(\lambda)$ which, consequently, are all real. They are also
positive, with the only possible exception of the lowest one,
according to the discussion in the previous section. Moreover, the zeroes of $f(\lambda)$ are simple and isolated; thus, the $\zeta$-function can be represented as the integral on the complex plane
\begin{equation}\label{z-rep}
  \zeta_\beta(s)=\frac{1}{2\pi
  i}\oint_{\mathcal{C}}\lambda^{-s}\frac{f'(\lambda)}{f(\lambda)}+
  \Theta(-\lambda_{0,\beta}) \lambda_{0,\beta}^{-s},
\end{equation}
where $\mathcal{C}$ is a curve which encloses counterclockwise the positive zeroes of
$f(\lambda)$, $\lambda_{0,\beta}$ is the lowest eigenvalue and $\Theta(\cdot)$ is the Heaviside function.

For $\Re(s)>1$ the path of integration in
(\ref{z-rep}) can be deformed to a vertical line to get
\begin{eqnarray}\label{z-rep-imag}
  \zeta_\beta(s)=-\frac{1}{2\pi
  i}\int_{-i\infty+0}^{i\infty+0}\lambda^{-s}
  \frac{f'(\lambda)}{f(\lambda)}\, d\lambda+
  h_1(s)=\\\mbox{}
  =  -\frac{e^{-i s \pi/2}}{2\pi}  \int_{1}^{\infty}
        \frac{f'(i\mu)}{f(i\mu)}\,
        \mu^{-s}\,d\mu
        -\frac{e^{i s \pi/2}}{2\pi}
        \int_{1}^{\infty} \frac{f'(-i\mu)}{f(-i\mu)}\,
        \mu^{-s}\,d\mu + h_2(s)\nonumber
\end{eqnarray}
where $h_1(s),h_2(s)$ are holomorphic functions. Eq.\ (\ref{z-rep-imag}) gives an integral representation of $\zeta_\beta(s)$, analytic in the half-plane $\Re(s)>1$. To compute its meromorphic extension to the whole complex $s$-plane and its pole structure we need the asymptotic expansion of
$f'(\lambda)/f(\lambda)$, which is given by \cite{FPW}
\begin{eqnarray}\label{te}\begin{array}{c}
    \displaystyle{
  \frac{f'(\lambda)}{f(\lambda)}\sim
    \frac{1}{4}\log{(-\lambda)}+
    \frac{1}{4}
    \sum_{k=0}^\infty c_k(\kappa)\,(-\lambda)^{-k}+ }\\
    \\
    \displaystyle{
    +\sum_{N=1}^\infty \sum_{n=0}^\infty
    C_{N,n}(\kappa,\beta)\,(-\lambda)^{-N(2\kappa-1)-2n-1}}\,.
\end{array}
\end{eqnarray}
The coefficients $c_k(\kappa)$ are polynomials in $\kappa$. As we will see, these terms do not contribute to the pole structure of $\zeta_\beta(s)$. On the other hand, the coefficients $C_{N,n}(\kappa,\beta)$ are defined through the following relations:
\begin{eqnarray}\label{CNn}
  C_{N,n}(\kappa,\beta)
  :=-\left( 4^{2\kappa-1} \beta \right)^N \left(
  {2\kappa-1+\frac{2n}{N} }\right) b_{n}(\kappa,N)\,,\\
  \sum_{n=0}^{\infty}b_n(\kappa,N)\,z^{-2n} :=
  \exp{\left\{N \sum_{m=1}^\infty a_m(\kappa)\,
  z^{-2m}\right\}}\,,\\
  a_m(\kappa):=\frac{2^{4m-1}}{2m+1}\left\{
  \left[ (1-\kappa)^{2m} - \kappa^{2m} \right]+
  \left(\frac{\kappa-1/2}{m}\right) \times\right.\\
  \times\left[ (1-\kappa)^{2m} + \kappa^{2m}  \right]+
  (2m+1) \sum_{p=1}^m \frac{B_{2p}}{p(2p-1)} \times
  \nonumber\\
  \left. \times
  \left(\begin{array}{c}
    2m-1 \\
    2p-2
  \end{array}\right) \left[ \kappa^{2(m-p)+1} -
  (1-\kappa)^{2(m-p)+1} \right] \right\}\,.\nonumber
\end{eqnarray}
Notice $C_{N,n}(\kappa,\beta)=0$ for $\beta=0$.

Replacing into Eq.\ (\ref{z-rep-imag}) the dominant logarithmic term in Eq.\ (\ref{te}) we get a simple pole at $s=1$ with residue $1/4$. The remaining terms in the asymptotic expansion of
$f'(\lambda)/f(\lambda)$ are of the form $A_j (-\lambda)^{-j}$,
for some $j\geq 0$  (see Eq.\ (\ref{te})). Replacing these terms into Eq.\
(\ref{z-rep-imag}) gives simple poles at $s=1-j$, with residues given by
$\displaystyle{- (A_j/\pi)\, \sin(\pi j)}$. Notice that these residues vanish for integer values of $j$.

In conclusion, there is a  simple pole at $s=1$ with residue $1/4$ as for the SAE with $\beta=0,-\infty$. But for a general SAE $H_{(\beta)}$ with $\beta\neq 0,-\infty$ and for $\frac{3}{4}\leq \kappa<1$ there are also simple poles at non integer values of $s$. Indeed, for each pair of integers $(N,n)$ with $N=1,2,3,\ldots$ and $n=0,1,2,\ldots$ the function $\zeta_\beta(s)$ has a simple pole at the negative value
\begin{equation}\label{pop}
    s_{N,n}=-N(2\kappa-1)-2n\,,
\end{equation}
with a $\beta$-dependent residue given by
\begin{equation}\label{res}
  \left. {\rm Res}\,\left[ \zeta_\beta(s) \right]\right|_{s=s_{N,n}} =
    \frac{(-1)^{N} }{\pi} \, C_{N,n}(\kappa,\beta)\, \sin(2\pi N\kappa)\,.
\end{equation}
Let us remark that when $\kappa$ is a rational number, there can
be several (but a finite number of) pairs $(N,n)$ contributing to
the same pole. On the contrary, when $\kappa$ is irrational the poles coming from
different pairs $(N,n)$ -which are also irrational- do not coincide.

Finally, notice that a pole of $\zeta_\beta(s)$ at a non
integer $s_{N,n}=-N(2\kappa-1)-2n$  implies the
presence of a term proportional to $t^{N(2\kappa-1)+2n}$ in the small-$t$ asymptotic expansion of
Tr$\displaystyle{\left\{e^{-t\, H_{(\beta)}}\right\}}$ (See Appendix \ref{spectralfunctionsrelations}).

Therefore, for this second order differential operator on the half-line we obtain similar results as in Section \ref{second-order} for a compact segment: There are two SAE whose definition domains are locally scale invariant near the singularity (See footnote \ref{dosLSInv}) and show the usual properties in their spectral functions. Moreover, there exists a continuous family of other SAE which presents anomalous poles in their $\zeta$-functions, dependent on an \emph{external parameter} (the coupling $\nu$), with residues dependent on the SAE.

\medskip

We finish this section by presenting an alternative method based on the relation between the pole structure of the $\zeta$-function and the asymptotic growth of the eigenvalues, which confirms our results. Indeed, the pole structure of $\zeta_\beta(s)$ can also be obtained from the asymptotic expansion of the eigenvalues
$\lambda_{n}$ for $n\gg 1$.

By solving Eq.\ (\ref{spe}) order by order in $n$ we get \cite{FPW}
\begin{eqnarray}\label{asymp-eigen}
   \lambda_{n}=4n+4( 1-\kappa) +
   \frac{4\beta }
  {\pi }\,\sin (2\,\pi \,\kappa )\, n^{1 - 2\,\kappa } +\\\mbox{}
  +
  \frac{4\beta }{
    \pi } \,\left( 1 - 3\,\kappa  + 2\,{\kappa }^2
      \right) \,\sin (2\,\pi \,\kappa )\, n^{-2\,\kappa }
    -\frac{ {2\beta }^2 }{\pi }\,
      \sin (4\,\pi \,\kappa )  \, n^{2 - 4\,\kappa }+\dots\,,\nonumber
\end{eqnarray}
where we have retained only powers of $n$ greater than $-2$. This
leads to
\begin{eqnarray}\label{asym-lambda}
  \zeta_\beta(s)
  = 4^{-s}\, \zeta(s)+s\, 4^{-s}\, (\kappa-1)\, \zeta( s+1)\, +\\\nonumber
  +  s\,
    \left(s + 1 \right)\, 4^{-s}
  \frac{{\left(\kappa -1  \right) }^2\, }{2}\,\zeta( s+2 )
  -s\,4^{-s}\, \frac{\beta }{\pi }\,
      \sin \left(  2\,\kappa  \,\pi\right) \,
      \zeta(s+2\,\kappa ) - \\\nonumber
      \mbox{}-s\,
      \left( s + 2\,\kappa  \right)4^{-s}\, \frac{\beta }{
      \pi } \,\left( \kappa  -1  \right)  \,
      \sin (2\,\pi \,\kappa )\,
      \zeta(1 + s + 2\,\kappa )\, + \\\nonumber
    + s\, 4^{-s}\,\frac{{\beta }^2}{2\,\pi } \,
    \sin (4\,\pi \,\kappa )\,
    \zeta( s-1  + 4\,\kappa ) \, +   \dots\,,
\end{eqnarray}
where $\zeta(z)$ is the Riemann $\zeta$-function. Taking into account that $\zeta(z)$ presents a
unique simple pole at $z=1$ with residue $1$, Eq.\ (\ref{asym-lambda}) confirms, order by order in this development, the pole structure we have already found.

\section{Krein Formula}\label{kf}

In this Section we study the behavior of the resolvent of a locally homogeneous second-order differential operator in relation with its selfadjoint extensions, and the consequences this has on the properties of other spectral functions. This will be done in the framework of the Krein's formula \cite{Krein1}, which relates the resolvent of two SAE of the given operator.

We consider the differential operator
\begin{equation}\label{mod}
    A=-\partial_x^2+\frac{\nu^2-1/4}{x^2}+V(x)\,,
\end{equation}
where $\nu\in(0,1)\subset\mathbb{R}$ and $V(x)$ is an analytic
function of $x\in\mathbb{R}^+$.

The operator (\ref{mod}) defined on $\mathcal{D}(A):=\mathcal{C}_0^\infty(\mathbb{R}^+)$ admits a continuous  family of selfadjoint extensions $A^\theta$ characterized by a real parameter which we call  $\theta$. Since the operator $e^{-tA^\theta}$ corresponding to a general SAE $A^\theta$ is not trace class (notice that the base manifold $\mathbb{R}^+$ is non-compact), we will consider the trace of the difference $e^{-tA^\theta}-e^{-tA^\infty}$, where $A^\infty$ corresponds to the Friedrichs extension \cite{Reed-Simon}. We will show in Theorem \ref{elthm11} that this trace admits an asymptotic expansion given by
\begin{equation}\label{weshow}
    {\rm Tr}\left\{e^{-tA^\theta}-e^{-tA^\infty}\right\}\sim
    \sum_{n=0}^\infty a_n(\nu,V)\,t^{\frac n 2}
    +\sum_{N,n=1}^\infty b_{N,n}(\nu,V)\,\theta^N\,
    t^{\nu N+\frac n 2-\frac 1 2}\,.
\end{equation}
As we will see, the SDW coefficients $a_n(\nu,V),b_{N,n}(\nu,V)$ can be recursively computed for each given potential $V(x)$. Let us remark that the singular term in (\ref{mod}) not only contributes to the coefficients
$a_n(\nu,V)$ of the standard powers of $t$ but also leads to the presence of non-standard powers of $t$ whose exponents are not half-integers but depend on the ``external'' parameter $\nu$. We will also show that
these terms are absent only for the SAE with $\theta=0$ and $\theta=\infty$, which correspond to selfadjoint extensions characterized by scale invariant domains. In Section
\ref{eg} we will consider the case $V(x)=x^2$ to compare with our results in Section \ref{adjoint-H}.

Actually, the main content of this section is the derivation of a generalization of the Krein's formula to the case of these kind of operators with singular coefficients, from which the expansion (\ref{weshow}) is obtained as a byproduct. We will make use of the two particular SAE of the operator $A$, namely $A^0$ and $A^\infty$, for which the $\nu$ dependent powers of $t$ in (\ref{weshow}) are absent. The expansion in Eq.\  (\ref{weshow}) will come out as a consequence of the relation between the resolvents corresponding to an arbitrary SAE and to $A^\infty$. This relation  is called Krein's formula \cite{Krein1} and has already been established for the case of operators with regular coefficients. We will therefore extend this result to $A$ in (\ref{mod}) and then use this generalization to prove the asymptotic behavior (\ref{weshow}).

\subsection{The regular coefficients case}\label{regu}

In this section we state without proof two theorems valid for the case of differential operators with regular coefficients. Theorem \ref{k0} describes the selfadjoint extensions
of a symmetric operator in terms of the (regular) boundary values of the functions belonging to its domains  \cite{Alb-Pan}. The statement of theorem \ref{KreinFormula} is the Krein's formula (see \cite{Achiezer}), which relates the resolvents corresponding to different selfadjoint extensions.
\begin{thm}\label{k0}
Let $A$ be a symmetric operator densely defined on a subspace
$\mathcal{D}(A)$ of a Hilbert space $\mathcal{H}$, for which the
deficiency indices are equal $n_+=n_-=:n<\infty$. Then:
\begin{itemize}
\item There exist two surjective maps
$\Gamma_1,\Gamma_2:\mathcal{D}(A^\dagger)\rightarrow \mathbb{C}^n$
such that $\forall\ \phi,\psi\in\mathcal{D}(A^\dagger)$
\begin{equation}
    (\phi,A^\dagger\psi)_\mathcal{H}-(A^\dagger\phi,\psi)_\mathcal{H}=
    (\Gamma_1\phi,\Gamma_2\psi)_{\mathbb{C}^n}-
    (\Gamma_2\phi,\Gamma_1\psi)_{\mathbb{C}^n}\,,
\end{equation}
where $(\cdot,\cdot)_\mathcal{H}$ is the inner product in
$\mathcal{H}$ and $(\cdot,\cdot)_{\mathbb{C}^n}$ is the usual
inner product in $\mathbb{C}^n$.

\item The selfadjoint extensions $A^{(M,N)}$ of $A$ are
characterized by two matrices $M,N\in\mathbb{C}^{n\times n}$, such
that $M\cdot N^\dagger$ is hermitian and
$(M|N)\in\mathbb{C}^{n\times 2n}$ has rank $n$. The domain of
definition of $A^{(M,N)}$ is defined as
\begin{equation}\label{cc1}
    \mathcal{D}\left(A^{(M,N)}\right):=\{\phi\in\mathcal{D}(A^\dagger):
    M \Gamma_1\phi=N \Gamma_2\phi\}\,.
\end{equation}
\end{itemize}
\end{thm}

Since the restrictions of $\Gamma_1,\Gamma_2$ to ${\rm
Ker}(A^\dagger-\lambda)$ are invertible, we can
establish the following definitions:
\begin{defn}\label{d1}
\begin{eqnarray}
    \Gamma_1^{-1}(\lambda):= \left(\Gamma_1|_{{\rm Ker}(A^\dagger-\lambda)}
    \right)^{-1}\,,\\
    K(\lambda):= -\Gamma_2\cdot \Gamma_1^{-1}(\lambda)\,.
\end{eqnarray}
\end{defn}

Now we can write down the Krein's formula, which expresses the resolvent\linebreak
$\left(A^{(M,N)}-\lambda\right)^{-1}$ corresponding to an
arbitrary selfadjoint extension in terms of the resolvent
$\left(A^{(\mathbf{1},\mathbf{0})}-\lambda\right)^{-1}$
corresponding to the selfadjoint extension characterized by the
matrices $M=\mathbf{1}$ and $N=\mathbf{0}$ \cite{derkach}.
\begin{thm}[Krein's formula]\label{KreinFormula}
\begin{equation}\label{krein0}
    \left(A^{(M,N)}-\lambda\right)^{-1}=
    \left(A^{(\mathbf{1},\mathbf{0})}-\lambda\right)^{-1}+
    \Gamma_1^{-1}(\lambda)\cdot\frac{N}{
    (M+N\,K(\lambda))}
    \cdot\left(\Gamma_1^{-1}(\lambda^*)\right)^\dagger\,.
\end{equation}
\end{thm}

\medskip

{\small\noindent
\textbf{Example}: let us write down the Krein's formula for the case of the
one-dimensional second order differential operator
\begin{equation}\label{kreinope}
    A=-\partial_x^2+U(x)\,,
\end{equation}
defined on  $\mathcal{C}_0^\infty(\mathbb{R}^+) \subset\mathbf{L_2}(\mathbb{R}^+)$. We assume that the potential $U(x)\in \mathcal{C}(\mathbb{R}^+)$
is in the limit point case \cite{Reed-Simon} at infinity and in the limit circle case  \cite{Reed-Simon}  at $x=0$. This is the case if there exist $\delta,\epsilon>0$
and a positive differentiable function $f(x)\geq -U(x)$ such that
\begin{eqnarray}
    0\leq U(x)\leq \frac{3/4-\epsilon}{x^2}\qquad \forall
    x\in(0,\delta)\,,\\
    \int_1^\infty \frac{dx}{\sqrt{f(x)}}\ {\rm diverges}\,,\quad
    {\rm and}\ \frac{f'(x)}{f(x)^{3/2}}\ {\rm is\ bounded\ near}\ \infty\,.
\end{eqnarray}
Then, the deficiency indices of $A$ are $n_\pm=1$ \cite{Reed-Simon}. In relation to Theorem \ref{k0}, we define the boundary operators
$\Gamma_1,\Gamma_2$ as
\begin{eqnarray}
    \Gamma_1\phi(x):=\phi(0)\,,\\
    \Gamma_2\phi(x):=\phi'(0)\,.
\end{eqnarray}
According to the second statement in Theorem \ref{k0}, the
selfadjoint extensions $A^\theta$ of $A$ are characterized by a
real parameter $\theta$ corresponding to $M^{-1}\,N\in \mathbb{R}$ and their
domains of definition $\mathcal{D}(A^\theta)$ are given by (see
Eq.\ (\ref{cc1}))
\begin{equation}\label{cc}
    \mathcal{D}\left(A^\theta\right)=\{\phi\in\mathcal{D}(A^\dagger):
    \phi'(0)-\theta\,\phi(0)=0\}\,.
\end{equation}
As expected, we obtain the classical boundary conditions of Robin
type. The extension characterized by $N=0$ (Dirichlet boundary
condition) corresponds to $\theta=\infty$ while $M=0$ (Neumann boundary conditions) corresponds to $\theta=0$. Notice that there
are only two boundary conditions, namely $\theta=0$ and $\theta=\infty$, which are scale invariant.

Let us now determine the operators $\Gamma_1^{-1}(\lambda)$ and
$K(\lambda)$ defined in (\ref{d1}). Since the deficiency
indices of the operator $A$ are $n_\pm=1$, the deficiency subspace
${\rm Ker}(A^\dagger-\lambda)$ is generated by a normalized
function which we denote by $\phi_\lambda (x)$. Consequently,
\begin{equation}\begin{array}{c}
    \Gamma_1^{-1}(\lambda):\mathbb{C}\rightarrow {\rm
    Ker}(A^\dagger-\lambda)\,,\\ \\
    \Gamma_1^{-1}(\lambda)\cdot c=
    \phi_\lambda(x)/\phi_\lambda(0)\cdot c\,.
\end{array}
\end{equation}
The operator $K(\lambda)$ is therefore given by
\begin{equation}\label{fak}
    K(\lambda)=-\frac{\phi_{\lambda}'(0)}{\phi_\lambda(0)}\,.
\end{equation}
The Krein's formula (Eq. (\ref{krein0})) can then be written as
\begin{equation}\label{krein}
    \left(A^{\theta}-\lambda\right)^{-1}-
    \left(A^{\infty}-\lambda\right)^{-1}=
    \frac{\left(A^{0}-\lambda\right)^{-1}-
    \left(A^{\infty}-\lambda\right)^{-1}}{1+\theta\, K(\lambda)}\,.
\end{equation}

Equation (\ref{krein}) gives the resolvent corresponding to an
arbitrary selfadjoint extension $A^{\theta}$ in terms of the
resolvents corresponding to the boundary conditions which are scale
invariant, namely $\theta=0$ (Neumann) and $\theta=\infty$
(Dirichlet).
}

\medskip

Following \cite{Falomir:2005xh}, in the next section we will prove a generalization of
the Krein's formula that relates in a similar way the resolvents corresponding to
different selfadjoint extensions of the operator in (\ref{mod}). The method employed follows the lines
given by E.\ Mooers in \cite{Mooers}. We will obtain an expression
similar to (\ref{krein})  in which the
$K(\lambda)$ factor, although not given by (\ref{fak}) as in the regular case, is also
related to the behavior near  the origin of the functions in ${\rm
Ker}\left(A^\dagger-\lambda\right)$.

\subsection{Locally homogeneous second order differential operators}\label{singu} \label{sec1}

So, we consider the symmetric differential operator $A$
\begin{equation}\label{sing}
    A=-\partial^2_x+\frac{\nu^2-1/4}{x^2}+V(x)\,,
\end{equation}
defined on $\mathcal{C}^\infty_0(\mathbb{R^+})\subset\mathbf{L_2}(\mathbb{R^+})$. We assume that $V(x)$ is an analytic function of $x\in\mathbb{R^+}$ bounded from below and the parameter $\nu\in(0,1)\subset\mathbb{R}$.

The following theorem describes the behavior of the functions in $\mathcal{D}(A^\dagger)$ near the singular point $x=0$:
{\thm If $\psi\in\mathcal{D}(A^{\dagger})$ then
\begin{equation}\label{comenelorieq}
    \psi(x)=C[\psi]\,\left(
    x^{-\nu+1/2}+
    \theta_{\psi}\,x^{\nu+1/2}\right)+O(x^{3/2})\,,
\end{equation}
for $x\rightarrow 0^+$ and some constants
$C[\psi],\theta_\psi\in\mathbb{C}$.}\label{comenelori}

\bigskip

{\noindent\bf Proof:} By virtue of Riesz representation lemma
\begin{equation}
\psi\in\mathcal{D}(A^{\dagger})\rightarrow
\exists\,\tilde{\psi}\in\mathbf{L_2}(\mathbb{R^+})\,:(\psi,A\phi)=(\tilde{\psi},\phi)\quad
\forall \phi\in\mathcal{D}(A)\,.
\end{equation}
Consequently,
\begin{equation}
A^{\dagger}\psi:=\tilde{\psi}\,.
\end{equation}
If we define $\chi:=x^{-\nu-1/2}\psi$ we obtain
\begin{equation}
    \partial_x(x^{2\nu+1}\partial_x\chi)=-x^{\nu+1/2}(\tilde{\psi}-V(x)\psi)
    \in\mathbf{L}_1(\mathbb{R^+})\,.
\end{equation}
Therefore, there exists a constant $C_1\in\mathbb{R}$ such that
\begin{equation}
    \partial_x\chi=C_1 x^{-1-2\nu}-x^{-1-2\nu}\int_0^x
    y^{\nu+1/2}\left(-\partial_y^2+\frac{\nu^2-1/4}{y^2}\right)\psi\,dy\,.
\end{equation}
The Cauchy-Schwartz inequality implies
\begin{eqnarray}
    \left|\,x^{-1-2\nu}\int_0^x y^{\nu+1/2}\left(-\partial_y^2+
    \frac{\nu^2-1/4}{y^2}\right)\psi\,dy\,\right|\leq\nonumber\\
    \leq C_2\,\left\|\left(-\partial_y^2+
    \frac{\nu^2-1/4}{y^2}\right)\psi\right\|_{(0,x)}\,x^{-\nu}\,,
\end{eqnarray}
for some $C_2\in\mathbb{R}$. In consequence,
\begin{eqnarray}
    \left|\,\int^x z^{-1-2\nu}\int_0^z
    y^{\nu+1/2}\left(-\partial_x^2+
    \frac{\nu^2-1/4}{x^2}\right)\psi\,dy\,dz\,\right|\leq\nonumber\\
    \leq C_3+C_4\,x^{1-\nu}\,,
\end{eqnarray}
where $C_3,C_4\in\mathbb{R}$. Thus, there exist
$C_5,C_6\in\mathbb{R}$, such that
\begin{equation}
    \psi=C_5\,x^{-\nu+1/2}+C_6\,x^{\nu+1/2}+O(x^{3/2})\,,
\end{equation}
for $x\rightarrow 0^+$.\fin

{\cor \label{Corooo}
\begin{equation}\label{corolario}
    \phi,\psi\in\mathcal{D}(A^{\dagger})\rightarrow
    (\phi,A^{\dagger}\psi)-(A^{\dagger}\phi,\psi)=
    C^*[\phi]C[\psi]\left(
    \theta^*_{\phi}-\theta_{\psi}\right)\,.
\end{equation}}

\bigskip

{\noindent\bf Remark 1:} Notice that (\ref{corolario}) verifies
the first statement of Theorem \ref{k0} according to the
definitions: $\Gamma_1\,\psi:=C[\psi]\,\theta_\psi$ and
$\Gamma_2\,\psi:=1$.

\bigskip

{\noindent\bf Remark 2:} By writing expression (\ref{corolario})
for $\psi=\phi$ we conclude that for all
$\psi\in\mathcal{D}(A^\dagger)$ the parameter $\theta_\psi$
defined by Theorem \ref{comenelori} is real.

\bigskip

{\noindent\bf Proof:} Expression (\ref{corolario}) follows from an
integration by parts in its {\small L.H.S.}\ using expression
(\ref{comenelorieq}).\fin

As a consequence of Corollary \ref{Corooo}, the differential
operator $A$ admits a family of selfadjoint extensions
$A^{\theta}$, characterized by the real parameter $\theta$, whose
domains are given by
\begin{equation}\label{saesing}
    \mathcal{D}(A^{\theta}):=
    \left\{\phi\in\mathcal{D}(A^{\dagger}):\theta_{\phi}=
    \theta\right\}\,,
\end{equation}
where $\theta_\phi$ is defined according to Theorem
\ref{comenelori}. The parameter $\theta$, with dimensions $[{\rm length}]^{-2\nu}$, thus determines the
\emph{boundary condition at the singularity}.

There exists another selfadjoint extension, which we denote by
$A^{\infty}$, whose domain is given by,
\begin{equation}\label{beta}
    \mathcal{D}(A^{\infty})=
    \left\{\phi\in\mathcal{D}(A^{\dagger}):
    \phi(x)=C[\phi]\,x^{\nu+1/2}+O(x^{3/2})\,,\ {\rm with\ }C[\phi]\in
    \mathbb{C}\right\}\,.
\end{equation}

Let us point out that in the regular case limit, when  $\nu\rightarrow 1/2$,
where the singular coefficient in the operator vanishes, this
parameter $\theta$ coincides with the one characterizing Robin
boundary conditions for the regular case (see Eq.\ (\ref{cc})).

\subsection{Generalization of the Krein's formula.}\label{sec2}

Our purpose now is to establish a relation between the resolvents
corresponding to different selfadjoint extensions of $A$.
This relation will prove to be useful to show that the trace of the
heat-kernel ${\rm Tr}\,e^{-tA^\theta}$ corresponding to a general
selfadjoint extension admits, for $\theta\neq 0,\infty$, a small-$t$
asymptotic expansion with $\nu$-dependent  powers of $t$.

We begin by stating the following theorem:
\begin{thm}\label{uni}
For any $f(x)\in\mathbf{L_2}(\mathbb{R}^+)$ and
$\lambda\notin\sigma(A^\theta)$ there exists a unique function
$\phi^{\theta}(x,\lambda)\in\mathcal{D}(A^{\theta})$ such that
\begin{equation}\label{pro}
    (A^{\theta}-\lambda)\,\phi^{\theta}(x,\lambda)
    =f(x)\,.
\end{equation}
Moreover,
\begin{equation}\label{inter}
    \phi^{\theta}(x,\lambda)=
    \int_0^{\infty}G_{\theta}(x,x',\lambda)f(x')\,dx'\,,
\end{equation}
being $G_{\theta}(x,x',\lambda)$ the kernel of the resolvent
$(A^{\theta}-\lambda)^{-1}$.
\end{thm}

The kernel $G_{\theta}(x,x',\lambda)$ can be written as
\begin{equation}\label{solres}
    G_\theta(x,x',\lambda)=-\frac{\Theta(x'-x)L_\theta(x,\lambda)R(x',\lambda)+
    \Theta(x-x')L_\theta(x',\lambda)R(x,\lambda)}{W[L_\theta,R](\lambda)}\,,
\end{equation}
where $\Theta(\cdot)$ is the Heaviside function. The functions $L_\theta(x,\lambda),R(x,\lambda)$ satisfy equation
(\ref{pro}) for $f(x)\equiv 0$. The latter is square integrable at
$x\rightarrow \infty$ and the former satisfies the boundary
condition
\begin{equation}\label{latac}
    L_\theta(x,\lambda)=x^{-\nu+1/2}+\theta\,x^{\nu+1/2}+O(x^{3/2})\,,
\end{equation}
at $x\rightarrow 0^+$. $W[L_\theta,R](\lambda)$ is the Wronskian
of $L_\theta(x,\lambda)$ and $R(x,\lambda)$, and is independent of
$x$.

To obtain the generalization of the Krein's formula we  begin by
relating the resolvents corresponding to $\theta=\infty$ and
$\theta=0$. In particular, for these selfadjoint extensions the
boundary condition (\ref{latac}) reads
\begin{equation}
    L_\infty(x,\lambda)=x^{\nu+1/2}+O(x^{3/2})\,,
\end{equation}
and
\begin{equation}
    L_0(x,\lambda)=x^{-\nu+1/2}+O(x^{3/2})\,.
\end{equation}
Since these functions determine the behavior at the origin of the
kernels $G_{\infty}(x,x',\lambda)$ and $G_{0}(x,x',\lambda)$, the
following definitions are in order:
\begin{defn}
\begin{eqnarray}
    G_{\infty}(x',\lambda):=\lim_{x\rightarrow
    0}x^{-\nu-1/2}\,G_{\infty}(x,x',\lambda)\,,\label{Rxp}\\
    G_{0}(x',\lambda):=\lim_{x\rightarrow
    0}x^{\nu-1/2}\,G_{0}(x,x',\lambda)\,.
\end{eqnarray}
\end{defn}

These new functions $G_{\infty}(x,\lambda)$ and $G_{0}(x,\lambda)$
determine the behavior at the singularity of the solutions
$\phi^{\infty}(x,\lambda)$ and $\phi^0(x,\lambda)$ of (\ref{pro})
corresponding to $\theta=\infty$ and $\theta=0$, respectively.
Indeed,
\begin{eqnarray}
    \ \ \ \phi^{\infty}(x,\lambda)=
    \int_0^{\infty}G_{\infty}(x,x',\lambda)f(x')\,dx'=
    \phi^{\infty}(\lambda)\,x^{\nu+1/2}+O(x^{3/2})\,,\label{fii}\\
    \phi^{0}(x,\lambda)=
    \int_0^{\infty}G_{0}(x,x',\lambda)f(x')\,dx'=
    \phi^{0}(\lambda)\,x^{-\nu+1/2}+O(x^{3/2})\,,\label{fi0}
\end{eqnarray}
being
\begin{eqnarray}
    \phi^{\infty}(\lambda):=
    \int_0^{\infty}G_{\infty}(x',\lambda)f(x')\,dx'\,,\label{fiienelori}\\
    \phi^{0}(\lambda):=
    \int_0^{\infty}G_{0}(x',\lambda)f(x')\,dx'\,.\label{fi0enelori}
\end{eqnarray}

To obtain a relationship between the kernels
$G_{\infty}(x,x',\lambda)$ and $G_{0}(x,x',\lambda)$ we will
relate the solutions $\phi^{\infty}(x,\lambda)$ and
$\phi^0(x,\lambda)$ of (\ref{pro}) corresponding to the same
inhomogeneity $f(x)$. To do that we need the following two
lemmas:

\begin{lem}\label{lem1}
Let $\varphi_0(x)\in\mathcal{D}(A^{0})$ such that
$\varphi_0(x)=x^{-\nu+1/2}+O(x^{3/2})$ for $x\rightarrow 0^+$.
Then, the solutions $\phi^{\infty}(x,\lambda)$ and
$\phi^0(x,\lambda)$ of (\ref{pro}) are related by
\begin{equation}\label{teo}
    \phi^{\infty}(x,\lambda)=\phi^{0}(x,\lambda)
    -\phi^{0}(\lambda)
    \left[\ \varphi_0(x)
    -\int_0^{\infty}G_{\infty}(x,x',\lambda)
    (A^0-\lambda)\varphi_0(x')\,dx'\ \right]\,.
\end{equation}
\end{lem}

\bigskip

\noindent{\bf Proof:} On the one hand,
\begin{equation}
    (A^\dagger-\lambda)\,\phi^\infty(x,\lambda)=f(x)\,.
\end{equation}
Moreover,
\begin{eqnarray}
    (A^\dagger-\lambda)\,\left\{
    \phi^{0}(x,\lambda)
    -\phi^{0}(\lambda)
    \left[\varphi_0(x)
    -\int_0^{\infty}G_{\infty}(x,x',\lambda)
    (A^0-\lambda)\varphi_0(x')\,dx'\right]
    \right\}=\nonumber\\=
    f(x)-\phi^{0}(\lambda)\left[(A^0-\lambda)\varphi_0(x)
    -(A^0-\lambda)\varphi_0(x)\right]=f(x)\,.\nonumber\\
\end{eqnarray}
On the other hand, for $x\rightarrow 0^+$,
\begin{equation}
    \phi^\infty(x,\lambda)=\phi^{\infty}(\lambda)\,x^{\nu+1/2}+O(x^{3/2})\,,
\end{equation}
and
\begin{eqnarray}
    \left\{
    \phi^{0}(x,\lambda)
    -\phi^{0}(\lambda)
    \left[\varphi_0(x)
    -\int_0^{\infty}G_{\infty}(x,x',\lambda)
    (A^0-\lambda)\varphi_0(x')\,dx'\right]
    \right\}=\nonumber\\=
    \phi^{0}(\lambda)
    \left[\int_0^{\infty}G_{\infty}(x',\lambda)
    (A^0-\lambda)\varphi_0(x')\,dx'\right]\cdot x^{\nu+1/2}
    +O(x^{3/2})
    \,.
    \nonumber\\
\end{eqnarray}
As a consequence, both sides of equation (\ref{teo}) belong to
$\mathcal{D}(A^\infty)$ and satisfy Eq.\  (\ref{pro}) for
$\theta=\infty$. Expression (\ref{teo}) is then proved by virtue
of the uniqueness stated in Theorem
\ref{uni}.\fin
\begin{lem}\label{lem2}
\begin{equation}
    \varphi_0(x)-\int_0^{\infty}G_{\infty}(x,x',\lambda)(A^0-\lambda)
    \varphi_0(x')\,dx'=
    2\nu\, G_{\infty}(x,\lambda)\,.
\end{equation}
\end{lem}

\noindent{\bf Proof:} Since the kernels of the resolvents
$G_{\infty}(x,x',\lambda),G_{0}(x,x',\lambda)$ are symmetric (see
Eq.\ (\ref{solres}))
\begin{equation}
    A^\dagger\,\left[G_{0}(x,x',\lambda)-G_{\infty}(x,x',\lambda)\right]=0\,,
\end{equation}
with $A^\dagger$ acting either on $x$ or $x'$. Thus,
\begin{eqnarray}
    \varphi_0(x)-\int_0^\infty G_\infty(x,x',\lambda)
    (A^0-\lambda)\varphi_0(x')\,
    dx'=\nonumber\\=
    \int_0^\infty\left[G_{0}(x,x',\lambda)-G_{\infty}(x,x',\lambda)\right]
    (A^0-\lambda)\varphi_0(x')\,dx'=\nonumber\\=
    \lim_{x\rightarrow 0^+}\left\{
    \left[G_{0}(x,x',\lambda)-G_{\infty}(x,x',\lambda)\right]\cdot
    \varphi_0'(x)-
    \right.\nonumber\\\left.\mbox{}-
    \partial_{x'}\left[G_{0}(x,x',\lambda)-G_{\infty}(x,x',\lambda)\right]\cdot
    \varphi_0(x)
    \right\}=\nonumber\\=
    \left[x^{-\nu+1/2}\,G_{0}(x,\lambda)-x^{\nu+1/2}\,G_{\infty}(x,\lambda)\right]
    \cdot(-\nu+1/2)\,x^{-\nu-1/2}-\nonumber\\
    \mbox{}-
    \left[(-\nu+1/2)\,x^{-\nu-1/2}\,G_{0}(x,\lambda)-
    (\nu+1/2)\,x^{\nu-1/2}\,G_{\infty}(x,\lambda)\right]\cdot
    x^{-\nu+1/2}=\nonumber\\=2\nu\,G_{\infty}(x,\lambda)\,.
\end{eqnarray}\fin
Lemmas \ref{lem1} and \ref{lem2} lead to the following result:
\begin{lem}\label{lemcon}
\begin{equation}\label{0vsinf}
    \phi^{0}(x,\lambda)=\phi^{\infty}(x,\lambda)+2\nu\, G_{\infty}
    (x,\lambda)\phi^{0}(\lambda)\,.
\end{equation}
\end{lem}

We are interested in rewriting expression (\ref{0vsinf}) so that
$\phi^{0}(x,\lambda)$ be given in terms of quantities
corresponding to the extension characterized by
$\theta=\infty$. In so doing, we take the $x\rightarrow 0^+$ limit  in equation
(\ref{0vsinf}) obtaining
\begin{equation}\label{Renelori}
    G_{\infty}(x,\lambda)=\frac{1}{2\nu}
    \left(x^{-\nu+1/2}-K(\lambda)^{-1}x^{\nu+1/2}\right)+O(x^{3/2})\,,
\end{equation}
where
\begin{equation}\label{Renelori2}
    K(\lambda):=\frac{\phi^{0}(\lambda)}{\phi^{\infty}(\lambda)}\,.
\end{equation}

The term $K(\lambda)$ defined in (\ref{Renelori2}) relates the
behavior at the singularity of the solutions to equation
(\ref{pro}) corresponding to the selfadjoint extensions
$\theta=\infty$ and $\theta=0$. Notice that Eq.\ (\ref{Renelori})
allows us to compute $K(\lambda)$ by studying the behavior at the
singularity of the kernel of the resolvent corresponding to the
extension $\theta=\infty$. Therefore, the kernel
$G_{\infty}(x,x',\lambda)$ determines $K(\lambda)$ and,
consequently, also $\phi^{0}(\lambda)$. Notice that $K(\lambda)$ reduces to
expression (\ref{fak}) for the regular limit $\nu\rightarrow 1/2$.

We can finally express the solution $\phi^{0}(x,\lambda)$ to
(\ref{pro}) corresponding to $\theta=0$ by means of the data
obtained by imposing the boundary conditions corresponding to
$\theta=\infty$ (see Lema \ref{lemcon}),
\begin{equation}\label{comp}
    \phi^{0}(x,\lambda)=\phi^{\infty}(x,\lambda)+2\nu K(\lambda)\,
    G_{\infty}(x,\lambda)\phi^{\infty}(\lambda)\,.
\end{equation}
Since this equation is valid for any inhomogeneity $f(x)$, by
virtue of equations (\ref{fii}), (\ref{fi0}) and
(\ref{fiienelori}), we obtain the following theorem:

\begin{thm}\label{uno}
\begin{equation}\label{R0vsRinf}
    G_{0}(x,x',\lambda)=G_{\infty}(x,x',\lambda)+2\nu K(\lambda)\,
    G_{\infty}(x,\lambda)G_{\infty}(x',\lambda)\,.
\end{equation}
\end{thm}

Next, we will establish an expression similar to (\ref{R0vsRinf})
giving the resolvent for an arbitrary selfadjoint extension in
terms of data related to the boundary conditions corresponding to
$\theta=\infty$. The first step is to state the following lemma:
\begin{lem}\label{rem} The solution $\phi^{\theta}(x,\lambda)$ to
(\ref{pro}) is given by
\begin{equation}\label{fithe}
    \phi^{\theta}(x,\lambda)=\phi^{\infty}(x,\lambda)+2\nu
 \left(K(\lambda)^{-1}+\theta\right)^{-1}G_{\infty}(x,\lambda)
 \phi^{\infty}(\lambda)\,.
\end{equation}
\end{lem}

\noindent{\bf Proof:} By means of equation (\ref{0vsinf}) it is
immediate to show that the difference between both sides of
expression (\ref{fithe}) belongs to ${\rm
Ker}(A^\dagger-\lambda)$.

On the other hand, both sides of (\ref{fithe}) belong to
$\mathcal{D}(A^\theta)$ since the behavior of its {\small R.H.S.}\
at the singularity is given by (see eqs.\ (\ref{fii}),
(\ref{Renelori}) and (\ref{Renelori2}))
\begin{equation}
    \frac{\phi^0(\lambda)\phi^\infty(\lambda)}{\phi^\infty(\lambda)+
    \theta\,\phi^0(\lambda)}\,
    \left(x^{-\nu+1/2}+\theta\,x^{\nu+1/2}\right)+O(x^{3/2})\,.
\end{equation}
Once more, uniqueness established in Theorem \ref{uni} leads us to
equation (\ref{fithe}).\fin

From Lemma \ref{rem}, together with equations (\ref{inter}),
(\ref{fii}) and (\ref{fiienelori}), we straightforwardly get the
following theorem: {\thm[Generalization of Krein's formula's]
\begin{equation}\label{RthevsRinf}
    G_{\theta}(x,x',\lambda)=G_{\infty}(x,x',\lambda)+2\nu
    \left(K(\lambda)^{-1}+\theta\right)^{-1}G_{\infty}(x,\lambda)
    G_{\infty}(x',\lambda)\,.
\end{equation}}

Expressions (\ref{R0vsRinf}) and (\ref{RthevsRinf}) readily lead
to
\begin{equation}\label{mira}
 G_{\theta}(x,x',\lambda)-G_{\infty}(x,x',\lambda)=
 \frac{G_{\theta}(x,x',\lambda)-G_{\infty}(x,x',
\lambda)}
    {1+\theta\, K(\lambda)}\,.
\end{equation}
Therefore we obtain the following relation between the resolvents
of the different selfadjoint extensions:
\begin{equation}\label{kreinsing}
    \left(A^{\theta}-\lambda\right)^{-1}-
    \left(A^{\infty}-\lambda\right)^{-1}=
    \frac{\left(A^{0}-\lambda\right)^{-1}-
    \left(A^{\infty}-\lambda\right)^{-1}}{1+\theta\, K(\lambda)}\,.
\end{equation}
This expression formally coincides with the Krein's formula
(\ref{krein}), which is valid for regular operators. However, while the
factor $K(\lambda)$ in (\ref{krein}) is given by (\ref{fak}), in the singular case under study $K(\lambda)$ in (\ref{kreinsing})
corresponds to equation (\ref{Renelori2}). As already mentioned
(\ref{krein}) and (\ref{kreinsing}) coincide in
$\nu\rightarrow 1/2$ limit.

\bigskip

We resume our results  in the following theorem, which
will allow us to prove the non-standard behavior of the spectral functions of the SAE of the operator $A$ in (\ref{sing}):
\begin{thm}\label{elthmenun}
\begin{equation}\label{elthm}
{{\rm Tr}\left\{(A^{\theta}-\lambda)^{-1}-
    (A^{\infty}-\lambda)^{-1}\right\}
    =\frac{{\rm Tr}\left\{(A^{0}-\lambda)^{-1}
    -(A^{\infty}-\lambda)^{-1}\right\}}
    {1+\theta\, K(\lambda)}}\,.
\end{equation}
\end{thm}

In the next section we will show that the asymptotic expansion of
$K(\lambda)$ for large $|\lambda|$ presents powers of $\lambda$
whose exponents depend on the parameter $\nu$. This will finally lead to the
asymptotic series (\ref{weshow}).

\subsection{Asymptotic expansion of the resolvent}\label{nocom}

In this section we will make use of Theorem \ref{elthmenun} to obtain the large-$|\lambda|$ asymptotic expansion for the resolvent $(A-\lambda)^{-1}$
of an arbitrary selfadjoint extensions of the operator $A$ defined in Eq.\
(\ref{sing}). According to this theorem it suffices to
study the solutions to
\begin{equation}\label{asiecu}
    (A+z)\psi=0\,,
\end{equation}
satisfying the boundary conditions corresponding to $\theta=\infty$ and $\theta=0$. If we consider $\lambda$ in the negative real semi-axis we can take $z\in\mathbb{R}^+$. In particular, we will focuss on the behavior
of the solutions for large $z$.

Taking into account the scaling properties of the first two terms
in (\ref{sing}) it will be convenient to define a new variable
$y:=\sqrt{z}\,x\in\mathbb{R}^+$. The solution to equation
(\ref{asiecu}) can then be written as $\psi=\psi(\sqrt{z}x,z)$, being $\psi(\sqrt{z}x,z)$ a solution to
\begin{equation}\label{asiecucamvar}
    \left(-\partial^2_y+\frac{\nu^2-1/4}{y^2}+1+\frac 1 z
    \,V(y/\sqrt{z})\right)\psi(y,z)=0\,.
\end{equation}
We propose the following Ansatz
\begin{equation}
    \psi(y,z)=\phi(y)+\sum_{n=0}^{\infty}\psi_n(y)z^{-1-n/2}\,,
\end{equation}
to be consistent with the analytic series for the potential,
\begin{equation}\label{pottay}
    V(x)=\sum_{n=0}^{\infty}V_nx^{n}\,,
\end{equation}
where $V_n:= V^{(n)}(0)/n!$. Eq.\ (\ref{asiecucamvar}) can be now solved order by order in $z$. The solution to
(\ref{asiecucamvar}) which is square integrable at $y\rightarrow
\infty$ can be written as
\begin{equation}\label{tam2}
    R(y,z)=\sqrt{y}K_\nu(y)+\sum_{n=0}^{\infty}\psi_n(y)z^{-1-n/2}.
\end{equation}
where
\begin{eqnarray}\label{fin}
    \psi_n(y)=\\
    \mbox{}-y^{1/2}K_\nu(y)\int_{0}^{y}\left[V_ny'^{(n+1/2)}K_{\nu}(y')+
    \!\!\!\!\sum_{l+m=n-2}\!\!\!\!
    V_{l}y'^l\psi_{m}(y')\right]\sqrt{y'}I_\nu(y')\,dy' -\nonumber\\
    \mbox{}-y^{1/2}I_\nu(y)\int_{y}^{\infty}\left[V_ny'^{(n+1/2)}K_{\nu}(y')+
    \!\!\!\!\sum_{l+m=n-2}\!\!\!\!
    V_{l}y'^l\psi_{m}(y')\right]\sqrt{y'}K_\nu(y')\,dy'\,.\nonumber\
\end{eqnarray}
Therefore, the behavior of $R(y,z)$ at $y\rightarrow 0^+$ is given by
\begin{equation}\label{venelori}
    R(y,z)\simeq\frac{\Gamma(\nu)}{2^{1-\nu}}\ y^{-\nu+1/2}+
    \frac{\Gamma(-\nu)}{2^{1+\nu}}\
    H(z)\cdot y^{\nu+1/2}
    +\ldots
\end{equation}
where we have defined
\begin{eqnarray}\label{hache}
    H(z):= 1+\frac{2\sin(\pi \nu)}{\pi}
    \times\\
    \times\sum_{n=0}^{\infty}
    z^{-1-n/2}
    \int_{0}^{\infty}
    \left[V_ny^{n+1/2}K_{\nu}(y)+\!\!\!\!\sum_{l+m=n-2}\!\!\!\!
    V_{l}y^l\psi_{m}(y)\right]\sqrt{y}K_\nu(y)\,dy\,.\nonumber
\end{eqnarray}
It is important to notice that $H(z)$ admits a large-$z$ asymptotic expansion in half-integer powers of $z$.

Next, we find a relation between $K(z)$ in equation (\ref{elthm}) (defined in Eq.\ (\ref{Renelori2})) and $H(z)$. To obtain an expression for $K(z)$ we study the kernel of the resolvent
$G_{\infty}(x,x',z)$ which, for $x<x'$, is given by (see Eq.\
(\ref{solres})),
\begin{equation}
    G_{\infty}(x,x',z)=\left.-\frac{z^{-1/2}}{W[L,R](z)}L(y,z)R(y',z)
    \right|_{y=\sqrt{z}x,y'=\sqrt{z}x'}\,.
\end{equation}
The function $L(y,z)$ is a solution to (\ref{asiecucamvar}) whose leading
term at the origin is proportional to $y^{\nu+1/2}$. $W[L,R](z)$
is the Wronskian of $L(y,z)$ and $R(y,z)$, which is independent of $y$.

According to definition (\ref{Rxp})
\begin{equation}\label{307}
    G_{\infty}(x',z)=\left.-\frac{z^{-1/2}y^{-\nu-1/2}}{W[L,R](z)}L(y,z)R(y',z)
    \right|_{y=0,y'=\sqrt{z}x'}\,.
\end{equation}
Replacing (\ref{venelori}) into equation (\ref{307}) we obtain the
behavior of $G_{\infty}(x,z)$ for $x\rightarrow 0^+$,
\begin{eqnarray}\label{re2}
    G_{\infty}(x\simeq 0,z)\simeq-
    \frac{\left.z^{-1/2}y^{-\nu-1/2}L(y,z)\right|_{y=0}}
    {W[L,R](z)}\times\\\times\,
    \left[\frac{\Gamma(\nu)}{2^{1-\nu}}(\sqrt{z}x)^{-\nu+1/2}+
    \frac{\Gamma(-\nu)}{2^{1+\nu}}\
    H(z)\cdot (\sqrt{z}x)^{\nu+1/2}\right]
    +\ldots\nonumber
\end{eqnarray}
Comparing equations (\ref{Renelori}) and (\ref{re2}) we get a
relation between $K(z)$ and $H(z)$,
\begin{equation}\label{kyh}
    K(z)=4^{\nu}\frac{\Gamma(1+\nu)}{\Gamma(1-\nu)}\;z^{-\nu}H(z)^{-1}\,.
\end{equation}
Since $H(z)$ admits an asymptotic expansion  in half-integer powers
of $z$, the large-$z$ asymptotic expansion of $K(z)$ contains powers of $z$
whose exponents depend on the parameter $\nu$. Theorem \ref{elthmenun} shows that this powers are also present in the large-$z$ asymptotic expansion of the resolvent trace, which after equation (\ref{elthm}) can be written as
\begin{equation}\label{elthm1}
{{\rm Tr}\left\{(A^{\theta}+z)^{-1}-
    (A^{\infty}+z)^{-1}\right\}
    =\frac{{\rm Tr}\left\{(A^{0}+z)^{-1}-(A^{\infty}+z)^{-1}\right\}}
    { 1+4^{\nu}\displaystyle{ \frac{\Gamma(1+\nu)}{\Gamma(1-\nu)} }\;
    \theta\;z^{-\nu}H(z)^{-1} } } \,.
\end{equation}
The trace in the {\small R.H.S.}\ of equation (\ref{elthm1}) can be readily
obtained from equation (\ref{R0vsRinf})
\begin{equation}\label{r0yinf}
    {\rm Tr}\left\{(A^{0}+z)^{-1}-(A^{\infty}+z)^{-1}\right\}=2\nu
    K(z)\int_0^\infty
    G_{\infty}^2(x,z)\,dx\,.
\end{equation}
To evaluate this expression we compare equations (\ref{Renelori}) and (\ref{re2}) and we get
\begin{equation}
    -\frac{\left.z^{-1/2}y^{-\nu-1/2}L(y,z)\right|_{y=0}}
    {W[L,R](z)}=\frac{1}{2^{\nu}\nu\Gamma(\nu)}\sqrt{z}^{\nu-1/2}\,.
\end{equation}
Therefore, equation (\ref{307}) reads
\begin{equation}
    G_{\infty}(x',z)=\frac{1}{2^{\nu}\nu\Gamma(\nu)}\sqrt{z}^{\nu-1/2}
    R(\sqrt{z}x',z) \,.
\end{equation}
Replacing this equation, together with (\ref{kyh}), into
expression (\ref{r0yinf}) we obtain
\begin{equation}\label{tam}
    {\rm Tr}\left\{(A^{0}+z)^{-1}-(A^{\infty}+z)^{-1}\right\}=
    \frac{2H(z)^{-1}\,z^{-1/2}}{\Gamma(\nu)\Gamma(1-\nu)}
    \int_0^\infty
    R(\sqrt{z}x,z)^2\,dx\,.
\end{equation}
This shows that ${\rm
Tr}\left\{(A^{0}+z)^{-1}-(A^{\infty}+z)^{-1}\right\}$ admits an
asymptotic expansion in half-integer powers of $z$.

The following theorem summarizes equations (\ref{elthm1}) and (\ref{tam}) regarding the resolvent-trace. We also state the corresponding result as regards the heat-kernel trace, which is the inverse Laplace transform of the resolvent-trace.

\begin{thm}\label{elthm11}\mbox{}\\
\begin{itemize}
\item
The trace of the difference between the resolvents
$(A^{\theta}-\lambda)^{-1}$ and $(A^{\infty}-\lambda)^{-1}$ admits
an asymptotic expansion for large $|\lambda|$ given by,
\begin{eqnarray}\label{resuinf}
    {\rm Tr}\left\{(A^{\theta}-\lambda)^{-1}-(A^{\infty}-
    \lambda)^{-1}\right\}\sim
    \sum_{n=2}^\infty \alpha_n(\nu,V)\,\lambda^{-\frac n 2}+\nonumber\\
    +\sum_{N,n=1}^\infty \beta_{N,n}(\nu,V)\,\theta^N\,
    \lambda^{-\nu N-\frac n 2-\frac 1 2}\,.
\end{eqnarray}
The coefficients $\alpha_n(\nu,V),\beta_n(\nu,V)$ depend on the
parameter $\nu$ characterizing the singularity and are also determined
by the coefficients $V_n$ characterizing the analytic potential
$V(x)$ by means of equations (\ref{elthm1}), (\ref{tam}),
(\ref{tam2}), (\ref{fin}) and (\ref{hache}) with
$z=e^{i\pi}\lambda$.

\bigskip

\item
The trace of the difference $e^{-t A^{\theta}}-e^{-t A^{\infty}}$ admits
a small-$t$ asymptotic expansion given by,
\begin{eqnarray}\label{hk}
    \mbox{}\\\nonumber{\rm Tr}\left\{e^{-t A^{\theta}}-e^{-t A^{\infty}}\right\}\sim
    \sum_{n=2}^\infty \frac{\alpha_n(\nu,V)}{\Gamma(n/2)}
    \,t^{\frac n 2-1}
    +\sum_{N,n=1}^\infty
    \frac{\beta_{N,n}(\nu,V)}{\Gamma(\nu N+\frac n 2+\frac 1 2)}
    \,\theta^N\,
    t^{\nu N+\frac n 2-\frac 1 2}\,.
\end{eqnarray}
\end{itemize}
\end{thm}
\fin

Let us give a dimensional analysis argument to explain  the non-standard powers of $t$ in the expansion (\ref{hk}). First of all, notice that the parameter $\theta$ introduced by the boundary conditions has dimensions $[{\rm length}]^{-2\nu}$ and, after the analyticity of $V(x)$, the dimensions of every other parameter in the problem is an integer power of the length. Since $t$ has dimensions $[{\rm length}]^{2}$, if the coefficients of the asymptotic expansion of the heat-trace were to depend analytically on $\theta$, then it is necessary that this expansion contains  integer powers of $t^\nu$. The only selfadjoint extensions for which these powers are to be absent are those with $\theta=0$ and $\theta=\infty$.

\subsection*{Example:}\label{eg}

Let us consider  $V(x)=x^2$. We will use the expansion (\ref{hk}) to describe the pole structure of the difference between the $\zeta$-functions $\zeta_A^\theta(s)-\zeta_A^\infty(s)$ corresponding to the operator (\ref{sing}), which will confirm the results obtained in Section \ref{adjoint-H} with two other techniques.

First of all, notice that for $V(x)=x^2$ only one of the coefficients $V_n$
defined in (\ref{pottay}) is non vanishing, namely
\begin{equation}
    V_n=\delta_{n,2}\,.
\end{equation}
As a consequence, the only functions $\psi_n(y)$ which are non-trivial corresponds to $n=2+4k$ with $k=0,1,\ldots$ (see Eq.\ (\ref{fin})). According to equations (\ref{hache}) and (\ref{tam2}),
\begin{eqnarray}
    H(z)^{-1}\sim 1+\sum_{k=1}^{\infty}C_k(\nu)\,z^{-2k}\,,\label{hasim}\\
    R(y,z)\sim \sqrt{y}K_\nu(y)+\sum_{k=1}^{\infty}C'_k(\nu,y)\,z^{-2k}\,,
\end{eqnarray}
for some $C_k(\nu),C'_k(\nu,y)$. Substituting these
equations into (\ref{tam}) we get
\begin{eqnarray}\label{fina}
    {\rm Tr}\left\{(A^{0}+z)^{-1}-(A^{\infty}+z)^{-1}\right\}\sim
    \nu\,z^{-1}+\sum_{k=0}^{\infty}C''_k(\nu)\,z^{-3-2k}\,,
\end{eqnarray}
where $C''_k(\nu)$ can be written in terms of
$C_k(\nu),C'_k(\nu,y)$.

Replacing equations (\ref{fina}) and (\ref{hasim}) into
(\ref{elthm1}) we obtain the asymptotic
expansion of the trace of the difference between the resolvents
$(A^{\theta}+z)^{-1}$ and $(A^{\infty}+z)^{-1}$
\begin{eqnarray}\label{desarro}
    {\rm Tr}\left\{(A^{\theta}+z)^{-1}-(A^{\infty}+z)^{-1}\right\}
    \sim \left[\nu\,z^{-1}+
    \sum_{k=0}^{\infty}C''_k(\nu)\,z^{-3-2k}\right]\times\\
    \times
    \sum_{N=0}^{\infty}
    (-1)^N 4^{N\nu}\left[\frac{\Gamma(1+\nu)}
    {\Gamma(1-\nu)}\right]^N \theta^N\,z^{-N\nu}
    \left[1+\sum_{k=1}^{\infty}C_k(\nu)\,z^{-2k}\right]^N\,.\nonumber
\end{eqnarray}
It is straightforward to see that the first series in (\ref{desarro}) gives no contribution to the pole structure of the difference $\zeta_A^\theta(s)-\zeta_A^\infty(s)$. On the other hand, from the second series in the asymptotic expansion (\ref{desarro}) one shows that $\zeta_A^{\theta}(s)-\zeta_A^{\infty}(s)$ has simple poles which are located at
\begin{equation}\label{toconf}
    s_{N,n}=-N\nu-2n\quad
    {\rm with}\ N=1,2,\ldots\ {\rm and}\ n=0,1,\ldots
\end{equation}
This result coincides with Eq.\ (\ref{pop}). In particular, the leading term in (\ref{desarro}) leads to a simple pole at
\begin{equation}
    s_{1,0}=-\nu\,,
\end{equation}
whose residue is given by
\begin{equation}\label{alfin}
    \frac{4^{\nu}}{\Gamma^2(-\nu)}\,
    \theta\,,
\end{equation}
in agreement with the result quoted in Eq.\ (\ref{res}).

\section{Conclusions}\label{conclusions}

We have studied some symmetric non essentially self adjoint first and second order differential operators with a singular potential term with the same scaling dimension as the highest derivative term, characteristic which we have mentioned as \emph{local scale invariance} (at the singular point in the potential). For certain range of the ``external parameters'' that weights  the singular term \emph{(i.e.}, the coupling constants in the potential -- $g$ or $\nu$ throughout the paper) these operators admit a continuous family of selfadjoint extensions. Each selfadjoint extension describes a different physical system, with its own spectrum determined by the behavior of the functions belonging to its domain near the singularity. This is an essential point in our discussion for it implies the existence of infinitely many physically admissible \emph{boundary conditions} at the singularity, identified in our examples by an additional dimensionful parameter (which is not present in the expression of the differential operator).

This situation takes place when the adjoint of the differential operators considered above admits two different square integrable behaviors near the singularity for the functions in its domain of definition. Then, a symmetric extension will contain, in general, both possible behaviors and then it must incorporate an additional dimensionful parameter necessary to specify its domain. The only exceptions are those combinations which make the domain (locally) scale invariant.

Therefore, the spectral functions associated to a selfadjoint extension of these operators will, in general, depend on this additional dimensionful parameter (with a scaling dimension which depends on the external parameters in the singular potential term), opening the possibility of having, for example, non-standard powers of $t$ in the asymptotic expansion of the the heat-kernel trace and, consequently, non-standard poles in the associated $\zeta$-function. By non-standard we mean that these powers and poles are not determined by the dimension of the base manifold and the order of the differential operator only (as in the smooth coefficients case) but also depend on the external parameters that characterize the singularity.

Our results show that those selfadjoint extensions which break in this way the local scale invariance of the dominant scaling dimension terms in the differential operator present these non-standard poles in the associated $\zeta$-function as well as non-standard powers of $\lambda$ in the large-$|\lambda|$ asymptotic expansion of  the resolvent.  As a consequence, for second order differential operators one finds non-standard powers of $t$ in the small $t$-asymptotic expansion of the heat-kernel trace. Moreover, the Seeley-De Witt coefficients and the residues of the $\zeta$- and $\eta$- functions depend on the selfadjoint extension.

In fact, also the large-$n$ asymptotic behavior of the eigenvalues $\lambda_n$ of the selfadjoint extensions contains powers of $n$ which depend on these external parameters. One also expects non-standard singularities in the corresponding Green functions at coincident points. This issue is relevant for the definition of physical states and the regularization of the stress-tensor, for example.

These non-standard behavior of the spectral functions has been explicitly shown by solving some examples of first and second order non essentially selfadjoint differential operators on both compact and non-compact one-dimensional base manifold. We have also proved that this phenomenon is not affected by the introduction of arbitrary smooth potentials.

Let us finally mention that, in establishing these results, we have derived an extension of the Krein's formula which applies to this kind of differential operators defined on functions which, generically, do not have a regular behavior at the singularity. This formula (Eqs.\ (\ref{RthevsRinf}) and (\ref{kreinsing})) relates the resolvents of different selfadjoint extensions  and directly leads to the non-standard  small-$t$ asymptotic expansion of the heat-kernel trace we have found.

\bigskip
\noindent
\textbf{Acknowledgements}: We are very pleased and we feel very honored for the possibility of contributing to this special volume in honor of Prof. Stuart Dowker. This  work was supported in part by grants from CONICET (PIP 01787), ANPCyT (PICT 00909) and UNLP (Proy.~11/X492), Argentina.

\appendix

\section{Spectral functions and its relations}\label{spectralfunctionsrelations}

Given an elliptic differential operator $A$ in a manifold $M$ with a complete orthogonal set of eigenfunctions corresponding to the eigenvalues
$\{\lambda_n\}_{n\in\mathbf{N}}$, then the associated $\zeta$-function is defined as \cite{Seeley,Gilkey}
\begin{equation}
    {
    \zeta_A(s)={\rm Tr}\,A^{-s}=\sum_{n\in\mathbf{N}}\lambda_n^{-s}\,,
    }
\end{equation}
which is a convergent series for  $\mathcal{R}(s)$ large enough.

If $\left|\lambda_n\right|\rightarrow\infty$ fast enough when $n\rightarrow\infty$, then also exists the trace of the resolvent, given by
\begin{equation}
    {
    {\rm Tr}\,(A-\lambda)^{-1}=\sum_{n\in\mathbf{N}}
    \frac{1}{\lambda_n-\lambda}\,.
    }
\end{equation}

Moreover, if the set $\left\{\Re\left(\lambda_n\right)\right\}$ is bounded below and $\Re\left(\lambda_n\right) \rightarrow\infty$ for $n \rightarrow \infty$, then the trace of the associated heat-kernel is given by
\begin{equation}
    {
    {\rm Tr}\,e^{-tA}=\sum_{n\in\mathbf{N}}e^{-t\lambda_n}\,.
    }
\end{equation}

For positive definite operators we have for the Laplace transform
\begin{equation}
    {
    {\rm Tr}\,(A-\lambda)^{-1}=\int_0^{\infty}e^{t\lambda}\,
    {\rm Tr}\,e^{-tA}\,dt\,, \label{reshea}
    }
\end{equation}
for $\mathcal{R}(\lambda)< \lambda_n\, , \forall\, n$, and for the Mellin transform
\begin{equation}
    {
    {\rm Tr}\,A^{-s}=\frac{1}{\Gamma(s)}\int_0^{\infty}t^{s-1}\,
    {\rm Tr}\,e^{-tA}\,dt\,,\label{zethea}
    }
\end{equation}
for $\mathcal{R}(s)>m/d$, where $m$ is the dimension of the manifold and $d$ the order of the differential operator.

\smallskip

The $\zeta$-function singularities are related with the asymptotic expansion of ${\rm Tr}\,(A-\lambda)^{-1}$ for large $|\lambda|$ and with the asymptotic expansion of ${\rm Tr}\,e^{-tA}$ for small values of $t$.
Indeed, if $\zeta_A(s)$ has simple poles at $s=s_n\leq s_0$, for $n\in\mathbf{N}$, then  ${\rm Tr}\,(A-\lambda)^{-1}$ has an asymptotic expansion in  powers of the form $\lambda^{s_n-1}$, while ${\rm Tr}\,e^{-tA}$ admits an asymptotic expansion in powers of the form $t^{-s_n}$. In particular, under the hypothesis considered in \cite{Seeley}, these powers depend only on $d$ and $m$. Moreover, the coefficients of both expansions are determined by the residues of $\zeta_A(s)$ at the corresponding poles.

\smallskip

For example, if
\begin{equation}\label{heat-kernel-trace}
    {
{ {\rm Tr} \{e^{-t A}\}\sim \sum_{n=0}^\infty c_n(A)\,
t^{\displaystyle{-s_n}}\,,}
    }
\end{equation}
with $s_n \leq s_0$, then for $\Re(s)> s_0$ we have
\begin{equation}\label{Mellin-asymp}{
    {
      \begin{array}{c}
    \displaystyle{{\Gamma(s)}\zeta_A(s)=\sum_{n=0}^{n_0} c_n(A)
    \int_0^1 t^{s-1-s_n}\, dt\, +}\\ \\
    \displaystyle{
    \int_0^1 t^{s-1}\left({\rm Tr} \{e^{-t A}\}-
    \sum_{n=0}^{n_0} c_n(A)\,
    t^{-s_n}
    \right) dt + }\\ \\
    \displaystyle{+
    \int_1^\infty t^{s-1}\,{\rm Tr} \{e^{-t A}\}\, dt
    =\sum_{n=0}^{n_0} c_n(A)\,
    \left(\frac{1}{s-s_n}\right)
    + h(s)\, ,}
    \end{array}
    }}
\end{equation}
where $h(s)$ is analytic on the open half plane $\Re(s)>s_{n_0}$. Therefore, the residue of ${\Gamma(s)}\zeta_A(s)$ at $s_n$ is related with the coefficient $c_n(A)$ through the equality
\begin{equation}\label{coef-residuos}
    {
{   \left.{\rm Res}
  \left[{\Gamma(s)}\zeta_A(s)\right]\right|_{s=s_n}= {c_n(A)}
  }}\,.
\end{equation}
Since ${\Gamma(s)}$ has simple poles at $s=0,-1,-2,\dots$, the residues $\left.{\rm Res}
\left[\zeta_A(s)\right]\right|_{s_n}$ vanish when $s_n=0,-1,-2, \dots$. In  particular, $\zeta_A(s)$ is analytic in a neighborhood of the origin.

\medskip

Even for nonpositive elliptic differential operators $A$, the complex  $s$-power of $A$ is defined in  terms of the resolvent as \cite{Seeley}
\begin{equation}\label{zetres-1}
    {
    A^{-s}=- \frac{1}{2\,\pi\,i} \oint_{\mathcal{C}}
  {\lambda^{-s}} \,\left(A-\lambda\right)^{-1}
  \, d\lambda\, ,
  }
\end{equation}
where $\mathcal{C}$ is a curve enclosing anti-clockwise the eigenvalues of $A$. From this one gets
\begin{equation}\label{zetres}
    {
    {\rm Tr}\,A^{-s}=- \frac{1}{2\,\pi\,i} \oint_{\mathcal{C}}
  {\lambda^{-s}} \, {\rm Tr}\left(A-\lambda\right)^{-1}
  \, d\lambda\, .
  }
\end{equation}

\medskip

The resolvent $(A-\lambda)^{-1}$, the complex power $A^{-s}$ and the heat-kernel $e^{-tA}$ can be considered as integral operators characterized by its kernels,  $G(x,x',\lambda)$, $\zeta_A(x,x',s)$ and $K(x,x',t)$, defined for $\lambda\in\mathbf{C}\backslash\{\lambda_n\}_{n\in\mathbf{N}}$, $\mathcal{R}(s)$ sufficiently large and  $t> 0$ respectively. In this case, their traces are expressed as
\begin{equation}\label{trre}{
    {\rm Tr}\,(A-\lambda)^{-1}=\int_M G(x,x,\lambda)\,dx\,,}
\end{equation}
\begin{equation}\label{113}{
    \zeta_A(s):= {\rm Tr}\,A^{-s}=\int_M \zeta_A(x,x,s)\,dx}
\end{equation}
and
\begin{equation}\label{trhe}{
    {\rm Tr}\,e^{-tA}=\int_M K(x,x,t)\,dx\,.}
\end{equation}


\section{The Hankel expansion}\label{Hankel}

In this appendix we write down some of the asymptotic expansions of the Bessel functions that are used in Section \ref{Asymptotic-expansion} \cite{A-S}. For $|z|\rightarrow \infty$, with $\nu$ fixed and
$|\arg z|<\pi$, we have
\begin{equation}\label{hankel}
  J_\nu(z)\sim \left(\frac{2}{\pi\,z}\right)^{\frac 1 2}
  \left\{ P(\nu,z) \cos \chi(\nu,z)
  -Q(\nu,z) \sin \chi(\nu,z) \right\}\, ,
\end{equation}
where
\begin{eqnarray}\label{chi}
  \chi(\nu,z) = z- \left(\frac \nu 2 + \frac 1 4\right) \pi,
\\ \label{P}
  P(\nu,z) \sim \sum_{k=0}^\infty (-1)^k\langle \nu , 2k\rangle
  \,\frac{1}{\left(2 z\right)^{2 k}}\, ,
\\\label{Q}
  Q(\nu,z) \sim \sum_{k=0}^\infty (-1)^k\langle \nu , 2k+1\rangle
  \,\frac{1}{\left(2 z\right)^{2 k+1}}\, ,
\end{eqnarray}
and the coefficients
\begin{equation}\label{coef-hankel}
  \langle \nu , k\rangle=
  \frac{\Gamma\left(\frac 1 2 +\nu + k\right)}
  {k! \, \Gamma\left(\frac 1 2 +\nu - k\right)}
  =\langle -\nu , k\rangle
\end{equation}
are the Hankel symbols. Therefore, $P(-\nu,z)=P(\nu,z)$, $Q(-\nu,z)=Q(\nu,z)$ and
\begin{equation}\label{Jupper}
  J_\nu(z)\sim
  \frac{e^{\mp i z}\, e^{\pm i \pi \left(
  \frac \nu 2 + \frac 1 4 \right)}}{\sqrt{2\pi z}}
  \sum_{k=0}^\infty \langle \nu , k\rangle
  \, \left(\frac{\mp i}{2 z}\right)^k\, ,
\end{equation}
where the upper (lower) signs correspond to $z$ in the upper (lower) open  half-plane. In particular, the quotient
\begin{equation}\label{JsobreJup}
      \frac{J_{\frac 1 2 -g}(\lambda)}{J_{g-\frac 1 2}(\lambda)}
      \sim  e^{\pm i \pi \left(\frac 1 2
      -g\right)}\, ,
\end{equation}
for $\Im(\lambda)>0$ and $\Im(\lambda)<0$, respectively.

Similarly, the derivative of the Bessel function has the following
asymptotic expansion for $|\arg z|<\pi$,
\begin{equation}\label{deriv-asymp}
  J'_\nu(z) \sim -\frac{2}{\sqrt{2 \pi  z}}
  \left\{
  R(\nu,z) \sin \chi(\nu,z) + S(\nu,z) \cos\chi(\nu,z)
  \right\}\, ,
\end{equation}
where
\begin{eqnarray}\label{R}
  R(\nu,z) \sim \sum_{k=0}^\infty(-1)^k\,
  \frac{\nu^2 + (2k)^2-1/4}
  {\nu^2-(2k-1/2)^2}\,
  \frac{\langle\nu,2k\rangle
  }{\left(2 z\right)^{2 k}}\, ,
\\\label{S}
  S(\nu,z) \sim \sum_{k=0}^\infty(-1)^k \,
  \frac{\nu^2 + (2k+1)^2-1/4}
  {\nu^2-(2k+1-1/2)^2}\,
  \frac{\langle\nu,2k+1\rangle
  }{\left(2 z\right)^{2 k+1}}\, .
\end{eqnarray}
Then,
\begin{equation}\label{Jprima}
  J'_\nu(z)\sim \mp i\,
  \frac{e^{\mp i z}\, e^{\pm i \pi \left(
  \frac \nu 2 + \frac 1 4 \right)}}{\sqrt{2\pi z}}
  \left\{ R(\nu,z)
  \mp i \,S(\nu,z) \right\}\, ,
\end{equation}
where the upper sign is valid for $\Im(z)>0$ and the lower
sign for $\Im(z)<0$. From the following relation
\begin{equation}\label{RS}
  R(\nu,z) \pm i \,S(\nu,z)=
  P(\nu,z) \pm i \,Q(\nu,z)
  + T_\pm(\nu,z)\, ,
\end{equation}
with
\begin{equation}\label{T}
  T_\pm(\nu,z)\sim
  \sum_{k=1}^\infty
  (2k-1)\langle\nu,k-1\rangle
  \left(\frac{\pm i}{2z}\right)^{k}\,,
\end{equation}
we get
\begin{equation}\label{JprimasobreJ}
    \frac{J'_\nu(z)}{J_\nu(z)}
  \sim \mp  i \left\{
  1+ \frac{T_\mp(\nu,z)}{P(\nu,z)\mp i Q(\nu,z)}\right\}
  \sim \mp  i \left\{
  1\mp \frac{i}{2z}+ O\left(\frac 1 {z^2}\right)\right\}\, ,
\end{equation}
where the upper sign is valid for $\Im(z)>0$ and the lower
one for $\Im(z)<0$. Finally, since the Hankel symbols are even in $\nu$ (see Eq.\
(\ref{coef-hankel})) we have
\begin{equation}\label{JprimasobreJasymp}
  \frac{J'_\nu(z)}{J_\nu(z)}
  \sim \frac{J'_{-\nu}(z)}{J_{-\nu}(z)}\, .
\end{equation}

\section{Closure of $H$} \label{closure}

In Section \ref{adjoint-H} we omit the contributions to
the boundary condition in Eq.\ (\ref{bc}) of the functions in the domain of the closure of the operator $H$ defined in Eq.\ (\ref{ham}); in this appendix we will justify this procedure. Indeed, we will show that
if $\phi\in{\mathcal D}(\overline{H})$ then
\begin{equation}\label{orders}
    \phi(x)=o(x^{\alpha}) \quad {\rm and}\quad
    \phi'(x)=o(x^{\alpha-1})
\end{equation}
near the origin, for any $\alpha:=\nu+\frac{1}{2}<3/2$.

In order to determine the domain $\mathcal{D}(\overline{H})$ of the closure of $H$  we
must consider those Cauchy sequences $\{\varphi_n\}_{n\in
\mathbb{N}}$ in ${\mathcal D}(H)={\mathcal C}_0^\infty (\mathbb{R^+})$, such that $\{H\varphi_n\}_{n\in \mathbb{N}}$ are
also Cauchy sequences. Since the coefficients of $H$
are real, it will suffice to consider real functions. Throughout this section $\varphi:=\varphi_n - \varphi_m$ (with $n,m\in
\mathbb{N}$) so that $\varphi \rightarrow 0$ and $H\varphi
\rightarrow 0$ as $n,m \rightarrow \infty$.

In the following we write $g:=\nu^2-\frac{1}{4}$. Notice first that the scalar product
\begin{equation}\label{phi-Hphi}
  \left( \varphi,H\varphi \right)
  =  \int_0^\infty \left( \varphi'^2 + \frac{g}{x^2}\,
  \varphi^2 +
  x^2 \varphi^2 \right)\, dx  \leq ||\varphi|| \, ||H\varphi||
  \rightarrow 0
\end{equation}
as $n,m \rightarrow \infty$; $\|\cdot\|$ represents the usual norm in $\mathbf{L_2}(\mathbb{R}^+)$. Therefore, for $g > 0$, we conclude
that
\begin{equation}\label{C-seq}
  \{\varphi'_n(x)\}_{n\in \mathbb{N}},\
  \displaystyle{\left\{\frac{\varphi_n(x)}{x}\right\}_{n\in
    \mathbb{N}}} \  {\rm and} \ \{x\, \varphi_n(x)\}_{n\in \mathbb{N}}
\end{equation}
are also Cauchy sequences. We will now prove the following

{\lem Let $\{\varphi_n\}_{n\in \mathbb{N}}$ be a
Cauchy sequence in ${\mathcal D}(H)={\mathcal
C}_0^\infty(\mathbb{R^+})$ such that, for $g>0$, $1\leq a<2$ and
$g\neq(a^2 -1)/4$,
\begin{equation}
  \{H\varphi_n\}_{n\in \mathbb{N}},\
  \displaystyle{\left\{\frac{\varphi_n(x)}{x^a}\right\}_{n\in
  \mathbb{N}}}, \ {\rm and} \
  \displaystyle{\left\{\frac{\varphi_n'(x)}{x^{a-1}}\right\}_{n\in
  \mathbb{N}}}
\end{equation}
are also Cauchy sequences. Then,
\begin{equation}
  \displaystyle{\left\{\frac{\varphi_n(x)}{x^{1+{a}/{2}}}\right\}_{n\in
    \mathbb{N}}} \ {\rm and} \
    \displaystyle{\left\{\frac{\varphi_n'(x)}{x^{{a}/{2}}}\right\}_{n\in
    \mathbb{N}}}
\end{equation}
are Cauchy sequences, too.\label{lemap}}

\smallskip

\noindent {\bf Proof:} Taking into account that the sum of fundamental sequences is
also a Cauchy sequence, we see that
\begin{equation}\label{l3}
  \left(A\,\frac{\varphi(x)}{x^a}+ B\, \frac{\varphi'(x)}{x^{a-1}}\, ,
  H \varphi(x) \right) \rightarrow 0
\end{equation}
as $n,m\rightarrow \infty$, for any pair of real numbers $A$
and $B$. It is easily seen that appropriately choosing the coefficients $A,B$ (whenever $g=(a^2 -1)/4$) Eq.\ (\ref{l3}) proves the lemma.\fin

Let us now assume that $g$ is an irrational number. Then,
applying iteratively Lemma \ref{lemap} to the sequences (\ref{C-seq}) one can show that,
for any positive integer $k$,
\begin{equation}\label{l6}
  \displaystyle{\left\{\frac{\varphi_n(x)}{x^{2\left[ 1- \left(
  1/2 \right)^k \right]}}\right\}_{n\in
    \mathbb{N}}} \quad {\rm and} \quad
    \displaystyle{\left\{\frac{\varphi_n'(x)}{x^{2\left[ 1- \left(
  1/2 \right)^k \right]-1}}\right\}_{n\in
    \mathbb{N}}}
\end{equation}
are Cauchy sequences, too. One immediately  concludes that
$\left\{x^{-2+\varepsilon}{\varphi_n(x)}\right\}_{n\in
\mathbb{N}}$ and \linebreak $\left\{x^{-1+\varepsilon}{\varphi_n'(x)}\right\}_{n\in
\mathbb{N}}$ are also Cauchy sequences. If $g$ were a rational number, one could choose an
irrational $a\in(1,3/2)$ from which Lemma \ref{lemap} could also be iteratively applied
to arrive to the same conclusions.

In the following we will consider the behavior of the functions
near the origin. For any $\varepsilon>0$, we can write
\begin{equation}\label{1}
   x^{-\alpha} \, \varphi(x) =
  \int_0^x \left( y^{-\alpha}  \, \varphi(y)
  \right)'\, dy
  =\int_0^x y^{{-\alpha} +1-\varepsilon}\left\{ {-\alpha}
  \,   \frac{\varphi(y)}{y^{2-\varepsilon}} +
  \frac{\varphi'(y)}{y^{1-\varepsilon}} \right\} \, dy \,.
\end{equation}
Therefore, for $x\leq 1$, ${\alpha} < 3/2$ and $\varepsilon$ small
enough, we have
\begin{equation}\label{2}
    \left| x^{-\alpha}  \, \varphi(x) \right|\leq
  \left(
  \int_0^1 y^{2({-\alpha} +1-\varepsilon)}dy \right)^{1/2}
  \left\{ |{\alpha} |\left\|
  \frac{\varphi(y)}{y^{2-\varepsilon}} \right\|
  +
  \left\|
  \frac{\varphi'(y)}{y^{1-\varepsilon}} \right\|  \right\}
  \rightarrow 0\,,
\end{equation}
as ${n,m \rightarrow \infty}$. We conclude that the sequence $\{x^{-\alpha}  \,
\varphi_n(x)\}_{n\in\mathbb{N}}$, for ${\alpha}<3/2$, is uniformly
convergent in $[0,1]$ and its limit is a continuous function
vanishing at the origin, which we write as $x^{-\alpha}  \,\phi(x)$:
\begin{eqnarray}\label{3}
  \lim_{n\rightarrow\infty}\left( x^{-\alpha}
  \,\varphi_n(x) \right)=
  x^{-\alpha}  \,\phi(x)\,,\\\label{4}
  \lim_{x\rightarrow 0^+} \left(x^{-\alpha}
  \,\phi(x)\right)=0\,.
\end{eqnarray}
In particular, for ${\alpha}  = 0$ we have the uniform limit
\begin{equation}\label{4-5}
 \lim_{n\rightarrow\infty}\varphi_n(x) = \phi(x)\,,
\end{equation}
which coincides with the limit of this sequence in
$\mathbf{L_2}(\mathbb{R^+})$.

Similarly, we can write
\begin{eqnarray}\label{6}
  x^{{-\alpha} +1}\,\varphi'(x)=- \int_0^x y^{{-\alpha} +1}\,H\varphi(y)\,dy
  +  \\
  + \int_0^x y^{{-\alpha} +1-\varepsilon} \left\{
  ({-\alpha} +1)\,
  \frac{\varphi'(y)}{y^{1-\varepsilon}}+ g\,
  \frac{\varphi(y)}{y^{2-\varepsilon}} \right\}\, dy +  \int_0^x y^{{-\alpha} +2}\, y\, \varphi(y)\, dy\,.\nonumber
\end{eqnarray}
Therefore, for $ x \leq 1$,  ${\alpha} < 3/2$ and $\varepsilon$
sufficiently small, we have
\begin{eqnarray}\label{7}
  \displaystyle{\left| x^{{-\alpha} +1}\,\varphi'(x) \right| \leq
  \left( \int_0^1 y^{2({-\alpha} +1)}\, dy \right)^{1/2} \left\|
  H\varphi(y) \right\| + } \\ \nonumber
  \displaystyle{\left( \int_0^1 y^{2({-\alpha} +1-\varepsilon)}\,
  dy \right)^{1/2}
  \left\{ |{\alpha} -1| \left\| \frac{\varphi'(y)}
  {y^{1-\varepsilon}}
  \right\|+
  g \left\| \frac{\varphi(y)}{x^{2-\varepsilon}} \right\|
  \right\}+ }\\ \nonumber
  \displaystyle{+ \left( \int_0^1 y^{2({-\alpha} +2)}\, dy \right)^{1/2}
  \left\|
  y\,\varphi(y) \right\| \rightarrow 0\,,}
\end{eqnarray}
as ${n,m\rightarrow \infty}$. Consequently, the sequence
$\{x^{{-\alpha}+1}\,\varphi_n'(x)\}_{n\in\mathbb{N}}$, with
${\alpha}< 3/2$, is uniformly convergent in $[0,1]$ and its limit
is a continuous function vanishing at the origin, which we write
as $x^{{-\alpha} +1} \,\chi(x)$:
\begin{eqnarray}\label{8}
  \lim_{n\rightarrow\infty}\left(
  x^{{-\alpha} +1}\,\varphi_n'(x) \right)=
   x^{{-\alpha} +1} \,\chi(x)\,,
\\\label{9}
  \lim_{x\rightarrow 0^+} \left( x^{{-\alpha} +1}
  \,\chi(x)\right)=0\,.
\end{eqnarray}
In particular, for ${\alpha} =1$ we have the uniform limit
\begin{equation}\label{10}
  \lim_{n\rightarrow\infty}\varphi'_n(x) = \chi(x),
\end{equation}
which coincides with the limit of this sequence in
$\mathbf{L_2}(\mathbb{R^+})$. Let us now show that $\chi(x)=\phi'(x)$. Indeed, for $x\leq 1$,
we have
\begin{eqnarray}\label{11}
  \left| \phi(x)-\int_0^x \chi(y)\,dy \right|\leq
  \\ \nonumber
  \leq \left| \phi(x)-\varphi_n(x) \right| +
  \left| \int_0^x \left(\chi(y)-\varphi'_n(y)\right)\,dy
  \right|\leq \left| \phi(x)-\varphi_n(x) \right| + \left\|
   \chi-\varphi'_n \right\| \rightarrow
   0\,,
\end{eqnarray}
as ${n \rightarrow \infty}$. Then, $\phi(x)$ is a differentiable function whose first derivative
is $\chi(x)$. On the other hand, eqs.\ (\ref{4}) and (\ref{9}) imply that any $\phi\in{\mathcal D}(\overline{H})$ satisfies Eq.\ (\ref{orders}).


\end{document}